\documentclass[journal]{IEEEtran}
\usepackage{graphicx}
\usepackage{hyperref}
\usepackage{amsfonts}
\usepackage{amsmath}
\usepackage{booktabs}
\usepackage{algorithm}
\usepackage[noend]{algpseudocode}
\usepackage{xcolor}
\usepackage[normalem]{ulem}

\begin{document}
%
% paper title
% Titles are generally capitalized except for words such as a, an, and, as,
% at, but, by, for, in, nor, of, on, or, the, to and up, which are usually
% not capitalized unless they are the first or last word of the title.
% Linebreaks \\ can be used within to get better formatting as desired.
% Do not put math or special symbols in the title.
\title{Unsupervised Domain Adversarial Self-Calibration for Electromyography-based Gesture Recognition}
%
%
% author names and IEEE memberships
% note positions of commas and nonbreaking spaces ( ~ ) LaTeX will not break
% a structure at a ~ so this keeps an author's name from being broken across
% two lines.
% use \thanks{} to gain access to the first footnote area
% a separate \thanks must be used for each paragraph as LaTeX2e's \thanks
% was not built to handle multiple paragraphs
%

\author{Ulysse Côté-Allard, Gabriel Gagnon-Turcotte, Angkoon Phinyomark, \\Kyrre Glette, Erik Scheme$\dagger$, François Laviolette$\dagger$, and~Benoit Gosselin$\dagger$
%\thanks{M. Shell was with the Department
%of Electrical and Computer Engineering, Georgia Institute of Technology, Atlanta,
%GA, 30332 USA e-mail: (see http://www.michaelshell.org/contact.html).}% <-this % stops a space
%\thanks{J. Doe and J. Doe are with Anonymous University.}% <-this % stops a space
%\thanks{Manuscript received April 19, 2005; revised August 26, 2015.}
\thanks{$\dagger$ These authors share senior authorship}
}

\maketitle

% As a general rule, do not put math, special symbols or citations
% in the abstract or keywords.
\begin{abstract}
Surface electromyography (sEMG) provides an intuitive and non-invasive interface from which to control machines. However, preserving the myoelectric control system's performance over multiple days is challenging, due to the transient nature of the signals obtained with this recording technique. In practice, if the system is to remain usable, a time-consuming and periodic recalibration is necessary. In the case where the sEMG interface is employed every few days, the user might need to do this recalibration before every use. Thus, severely limiting the practicality of such a control method. Consequently, this paper proposes tackling the especially challenging task of unsupervised adaptation of sEMG signals, when multiple days have elapsed between each recording, by introducing Self-Calibrating Asynchronous Domain Adversarial Neural Network (SCADANN). SCADANN is compared with two state-of-the-art self-calibrating algorithms developed specifically for deep learning within the context of EMG-based gesture recognition and three state-of-the-art domain adversarial algorithms. The comparison is made both on an offline and a dynamic dataset (20 participants per dataset), using two different deep network architectures with two different input modalities (temporal-spatial descriptors and spectrograms). Overall, SCADANN is shown to substantially and systematically improves classification performances over no recalibration and obtains the highest average accuracy for all tested cases across all methods.
\end{abstract}

\begin{IEEEkeywords}
EMG, Myoelectric Control, Domain Adaptation, Self-Calibration, Domain Adversarial, Gesture Recognition.
\end{IEEEkeywords}

\maketitle
\section{Introduction}

Robots have become increasingly prominent in the lives of human beings. As a result, the way in which people interact with machines is constantly evolving towards a better synergy between human intention and machine action. The ease of transcribing intention into commands is highly dependent on the type of interface and its implementations~\cite{interface_robot_intuitiveness}. Within this context, muscle activity offers an attractive and intuitive way to perform gesture recognition as a guidance method~\cite{robotic_arm_iros, emg_survey}. Such activity can be recorded from surface electromyography (sEMG), a non-invasive technique widely adopted both for prosthetic control and in research as a way to seamlessly interact with machines~\cite{prosthethics_review_erik, robot_dance_journal_paper}. 
%sEMG signals are non-stationary, and represent the sum of subcutaneous motor action potentials generated through muscular contractions~\cite{emg_survey}.
Artificial intelligence can then be leveraged as the bridge between these biological signals and a robot's input guidance.

Current state-of-the-art algorithms in gesture recognition routinely achieve accuracies above 95\% for the classification of offline, within-day datasets~\cite{journal_paper_TL_ulysse, convNetInstantImageEMG}. However, many practical issues still need to be solved before implementing these types of algorithms into functional applications~\cite{prosthethics_review_erik, overview_emg_interface_2015}. Electrode shift and the transient nature of the sEMG signals are among the main obstacles to a robust and widespread implementation of real-time sEMG-based gesture recognition~\cite{prosthethics_review_erik}. In practice, this means that users of current myoelectric systems need to perform periodic recalibration of their device so as to retain their usability. To address the issue of real-time myoelectric control, researchers have proposed rejection-based methods where a gesture is predicted only when a sufficient level of certainty is achieved~\cite{erik_confidence_base_prediction, adaptive_increase_window_size}. While this type of method have been shown to increase online usability, they do not directly address the inherent decline in performance of the classifier over time. One way to tackle this challenge is to leverage transfer learning algorithms to periodically recalibrate the system with less data than normally required~\cite{transfer_learning_emg_adaptation, VR_long_term_dataset}. Though this reduces the burden placed on the user, said user will still need to periodically record new labeled data.

This work focuses on the problem of across-day sEMG-based gesture recognition both within an offline and dynamic setting. In particular, this work considers the situation where several days have elapsed between each recording session. Such a setting naturally arises when sEMG-based gesture recognition is used for video games, artistic performances or, simply, to control devices of sporadic use~\cite{erik_emg_game, robot_dance_journal_paper, robotic_arm_iros}. In contrast to within-day or even day-to-day adaptation, this work's setting is especially challenging as the change in the signal between two sessions is expected to be substantially greater and no intermediary data is available to bridge this gap. The goal is then for the classifier to be able to adapt over-time using the unlabeled data obtained from the myoelectric system. Such a problem can be framed within an unsupervised domain adaptation setting~\cite{dann_OG} where there exists an initial labeled dataset on which to train, but the classifier then has to adapt to unlabeled data from a different, but similar distribution.

 An additional difficulty of the setting considered in this work is that real-time myoelectric control imposes strict limitations in relation to the amount of temporal data which can be accumulated before each new prediction. The window's length requirement has a direct negative impact on the performance of classifiers~\cite{optimal_latency_real_time_EMG, adaptive_increase_window_size}. This is due to the fact that temporally neighboring segments most likely belong to the same class~\cite{lda_ann_self_recalibration, self_recalibration_majority_vote}. In other words, provided that predictions can be deferred, it should be possible to generate a classification algorithm with improved accuracy (compared to the real-time classifier) by looking at a wider temporal context of the data~\cite{adaptive_increase_window_size}. Consequently, one possibility to cope with electrode shift and the non-stationary nature of EMG signals for gesture recognition is for the classifier to self-calibrate using pseudo-labels generated from this improved classification scheme.
 The most natural way of performing this relabeling is using a majority vote around each classifier's prediction. Xiaolong et al.~\cite{self_recalibration_majority_vote} have shown that such a recalibration strategy significantly improves intra-day accuracy on an offline dataset for both amputees and able-bodied subjects (tested on the NinaPro DB2 and DB3 datasets~\cite{ninaProDB2_DB3}). However for real-time control, such a majority vote strategy will increase latency, as transitions between gestures inevitably take longer to be detected. Additionally, as the domain divergence, over multiple days, is expected to be substantially greater than within a single day, ignoring this gap before generating the pseudo-labels might negatively impact the self-recalibrated classifier. Finally, trying to re-label every segment, even when there is no clear gesture detected by the classifier, will necessarily introduce undesirable noise in the pseudo-labels.
 To address these issues, the main contribution of this paper is the introduction of SCADANN (for Self-Calibrating Asynchronous Domain Adversarial Neural Network), a deep learning-based algorithm, which leverages the domain adaptation setting and the unique properties of real-time myoelectric control for inter-day self-recalibration.

This paper is organized as follows. An overview of the related work is given in Section~\ref{related_work}. The datasets and the deep network architecture employed in this work is provided in Section~\ref{dataset_and_convNetArchitecture}. Section~\ref{domain_adaptation} presents the domain adaptation algorithm considered in this work, while Section~\ref{unsupervised_recalibration} thoroughly describes SCADANN alongside the two most popular sEMG-based unsupervised adaptation algorithms. Finally, these three algorithms are compared alongside the domain adaptation algorithms and the vanilla networks in Section~\ref{result_section} and their associated discussions are shown in Section~\ref{discussion}. 

\section{Related Work}
\label{related_work}

Myoelectric control systems naturally generate large amounts of unlabeled data. However, over time, due to the electrode shift and transient change in the signal, the data generated diverges from the one used for training the classifier. Huang et al.~\cite{svm_particle_adaptive_emg} proposes using this setting to update a modified Support Vector Machine, by replacing some of the key examples (referred to as representative particles (RP)) from the training set, with new unlabeled examples when they are sufficiently close (i.e. small distance within the feature space) to the RP. Other authors~\cite{adaptation_replacing_dataset_with_newer_close_examples} propose to periodically retrain a Linear Discriminant Analysis (LDA), by updating the training dataset itself. The idea is to replace the oldest examples with new, near ones. Such methods, however, are inherently restricted to single-day use as they rely on smooth and small signal drift to update the classifier. Additionally, these types of methods do not leverage the potentially large quantity of unlabeled data generated. In contrast, deep learning algorithms are well suited to scale to large amounts of data and were shown to be more robust to between-day signal drift than LDA, especially as the amount of training data increases~\cite{deep_learning_between_days}. Within the field of image recognition, deep learning-based unsupervised domain adaptation has been extensively studied. A popular approach to this problem is domain adversarial training popularized by DANN~\cite{dann_OG, DANN}. The idea behind DANN is to learn a feature representation which favors class separability of the labeled dataset, while simultaneously hindering domain separability (i.e. differentiation between the labeled and unlabeled examples). See Section~\ref{domain_adaptation} for details. Building on DANN, the VADA (for Virtual Adversarial Domain Adaptation) algorithm~\cite{dirt_T} proposes to also minimize the cluster assumption violations on the unlabeled dataset~\cite{cluster_assumption_semi_supervised} (i.e. decision boundary should avoid area of high data density). Another state-of-the-art algorithm, but this time for non-conservative unsupervised domain adaptation (i.e. the final model might not be good at classifying the original data), is DIRT-T (for Decision-boundary Iterative Refinement Training with a Teacher), which starting from the output of VADA, removes the labeled data and iteratively tries to continue minimizing the cluster assumption. A detailed explanation of DANN, VADA and DIRT-T is given in Section~\ref{domain_adaptation}. These three state-of-the-art domain adversarial algorithms achieve a two-digit accuracy increase on several difficult image recognition benchmarks~\cite{dirt_T} compared to the non-adapted deep network. This work thus proposes to test these algorithms on the challenging problem of multiple-day sEMG-based gesture recognition both within an offline and dynamic setting (see Section~\ref{result_section}).

\section{Datasets and Network's Architecture}
\label{dataset_and_convNetArchitecture}
This work employs the \textit{3DC Dataset}~\cite{3dcArmband} for architecture building and hyperparameter optimization and the \textit{Long-term 3DC Dataset}~\cite{VR_long_term_dataset} for training and testing the different algorithms considered. Both datasets were recorded using the \textit{3DC Armband}~\cite{3dcArmband}; a wireless, 10-channel, dry-electrode, 3D printed sEMG armband. The device samples data at 1000 Hz per channel, allowing to take advantage of the full spectra of sEMG signals~\cite{emg_200_vs_1000Hz}.

As stated in~\cite{3dcArmband, VR_long_term_dataset}, the data acquisition protocol of the 3DC Dataset and Long-term 3DC Dataset were approved by the Comités d’Éthique de la Recherche avec des êtres humains de l’Université Laval (approval number: 2017-0256 A-1/10-09-2018 and 2017-026 A2-R2/26-06-2019 respectively), and informed consent was obtained from all participants.

\subsection{Long-term 3DC Dataset}
\label{longterm3DC_dataset_section}
\begin{figure}[h!]
\begin{center}
\includegraphics[width=\linewidth]{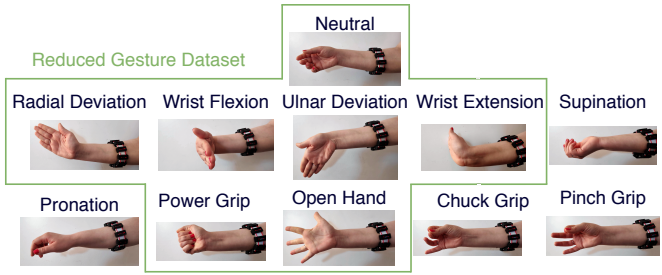}
\end{center}
\caption{The eleven hand/wrist gestures recorded in the \textit{Long-term 3DC dataset} and the \textit{3DC Dataset}. The gestures included within the \textit{Reduced Long-term 3DC Dataset} are encompassed within the green line (7 gestures totals).}
\label{fig_gestures_dataset}
\end{figure}

The Long-term 3DC Dataset features 20 able-bodied participants (5F/15M) aged between 18 and 34 years old (average 26$\pm$4 years old) performing eleven gestures (shown in Figure~\ref{fig_gestures_dataset}). Each participant performed three recording sessions over a period of fourteen days (in seven-day increments). Each recording session is divided into a \textit{Training Recording} and two \textit{Evaluation Recordings}. For each new session, the participants were the ones placing the armband on their forearm at the beginning of each session (introducing small electrode shift between each session).

The Long-term 3DC Dataset was recorded within a virtual reality environment in conjunction with the leap motion camera. The usefulness of the VR environment was three fold. First, it allowed to more intuitively communicate requested gesture intensity and position to the participant. Second, it allowed to replace the arm of the participant with a \textit{virtual prosthetic}, which provided direct and intuitive feedback (gesture held, intensity and position) to the participant. Third, it allowed the gamification of the experimental protocol, which greatly facilitated both recruitment and participant retention. During recording, the leap motion, in conjunction with an image-based convolutional network, served as the real-time controller and as a way to provide feedback without biasing the dataset to a particular EMG-based classifier.

The dataset is thoroughly described alongside a detailed explanation of the VR system and the contributions of the leap motion camera in~\cite{VR_long_term_dataset}. A brief overview of the dataset is provided in the following subsections. A video showing the recording protocol in action is also available at \href{https://www.youtube.com/watch?v=BnDwcw8ol6U}{the following link: https://www.youtube.com/watch?v=BnDwcw8ol6U}.

\subsubsection{Training Recording}
\label{trainingSession}
During the Training Recording, each participant was standing and held their forearm, unsupported, parallel to the floor, with their hand relaxed (neutral position). Starting from this neutral position, each participant was asked to perform and hold each gesture for a period of five seconds. This was referred to as a \textit{cycle}. Two more such cycles were recorded. In this work, the first two cycles are used for training, while the last one is used for testing (unless specified otherwise). Note that in the original dataset, four cycles are recorded for each participant, with the second one recording the participant performing each gesture with maximal intensity. This second cycle was removed for this work to reduce confounding factors. In other words, cycle two and three in this work correspond to cycle three and four in the original dataset.

In addition to the eleven gestures considered in the Long-term 3DC Dataset, a reduced dataset from the original Long-term Dataset containing seven gestures is also employed. This \textit{Reduced Long-term 3DC Dataset} is considered as it could more realistically be implemented on a real-world system given the current state of the art of EMG-based hand gesture recognition. The following gestures form the reduced dataset: neutral, open hand, power grip, radial/ulnar deviation and wrist flexion/extension. These gestures were selected as they were shown to be sufficient in conjunction with orientation data to control a 6 degree-of-freedom robotic arm in real-time~\cite{robotic_arm_iros}. 

\subsubsection{Evaluation Recording}
\label{evaluationSession}

During the Evaluation Recordings, the participants were asked to perform a specific gesture at a specific intensity (low, medium and high intensity based on their corresponding maximal gesture intensity) and at a random position (a point within reach of the participant's extended arm at a maximum angle of $\pm$45 and $\pm$70 degrees in pitch and yaw respectively). A new gesture, intensity and position were randomly asked every five seconds. Each Evaluation Recording lasted three and a half minutes and two such recordings were performed by each participant for each recording session (total of six Evaluation Recordings per participant). The Evaluation Recordings provide a dynamic dataset which includes the transitions between the different gestures and the four main dynamic factors~\cite{prosthethics_review_erik} (i.e. contraction intensity, inter-day recording, electrode shifts and limb position) in sEMG-based gesture recognition. Note that while the participants received visual feedback within the VR environment in relation to the held gesture, limb position and gesture intensity, the performed gestures were classified using the leap motion camera~\cite{leap_motion_patent} in order to avoid bias in the dataset towards a particular EMG-based classifier. In other words, the controller used by the participants during the Evaluation Recordings is distinct and independent from the sEMG-based gesture recognition algorithms considered in this manuscript, which is the main difference between the dynamic dataset considered and a real-time dataset. In this work, the first evaluation recording of a given session was employed as the unlabeled training dataset for the algorithms presented in Section~\ref{domain_adaptation} and~\ref{unsupervised_recalibration}, while the second evaluation recording was used for testing.

\subsubsection{Data Pre-processing}
\label{preprocessing_section}
This work aims at studying unsupervised recalibration of myoelectric control systems. Consequently, the input latency is a critical factor to consider. The optimal guidance latency was found to be between 150 and 250 ms~\cite{optimal_latency_real_time_EMG}. As such, the data from each participant is segmented into 150 ms frames with an overlap of 100 ms. Each segment thus contains $10\times150$ ($channel\times time$) data points. The segmented data is then band-pass filtered between 20-495 Hz using a fourth-order butterworth filter.

Given a segment, the spectrogram for each sEMG channel are then computed using a 48 points Hann window with an overlap of 14 yielding a matrix of $4\times25$ ($time\times frequency$). The first frequency band is then removed in an effort to reduce baseline drift and motion artifacts. Finally, following~\cite{transfer_learning_conference_emg}, the time and channel axis are swapped such that an example is of the shape $4\times 10 \times 24$ ($time \times channel \times frequency$). Spectrograms were selected as inputs for the ConvNet presented in Section~\ref{spectrogram_architecture}, as they have been shown to obtain competitive performance on a wide variety of datasets~\cite{journal_paper_TL_ulysse, self_recalibration_majority_vote, 3dcArmband} and in the control of a robotic arm in real-time~\cite{robotic_arm_iros}. In addition, they are relatively inexpensive to compute and allow for faster training of a ConvNet when compared to the raw sEMG signal due to the relatively low dimensionality of the obtained input images from the spectrograms.

\subsection{3DC Dataset}
The 3DC Dataset features 22 able-bodied participants and is employed for architecture building and hyperparameter selection. This dataset, presented in~\cite{3dcArmband}, includes the same eleven gestures as the Long-term 3DC Dataset. Its recording protocol closely matches the Training Recording description (Section~\ref{longterm3DC_dataset_section}), with the difference being that two such recordings were taken for each participant (within the same day). This dataset was preprocessed as described in Section~\ref{preprocessing_section}. %Note that when recording the 3DC Dataset, participants were wearing both the Myo and 3DC Armband, however in this work, only the data from the 3DC Armband was used.  

\subsection{Convolutional Network's Architecture}
\label{spectrogram_architecture}
\begin{figure*}[!htbp]
\centering
\includegraphics[width=\textwidth]{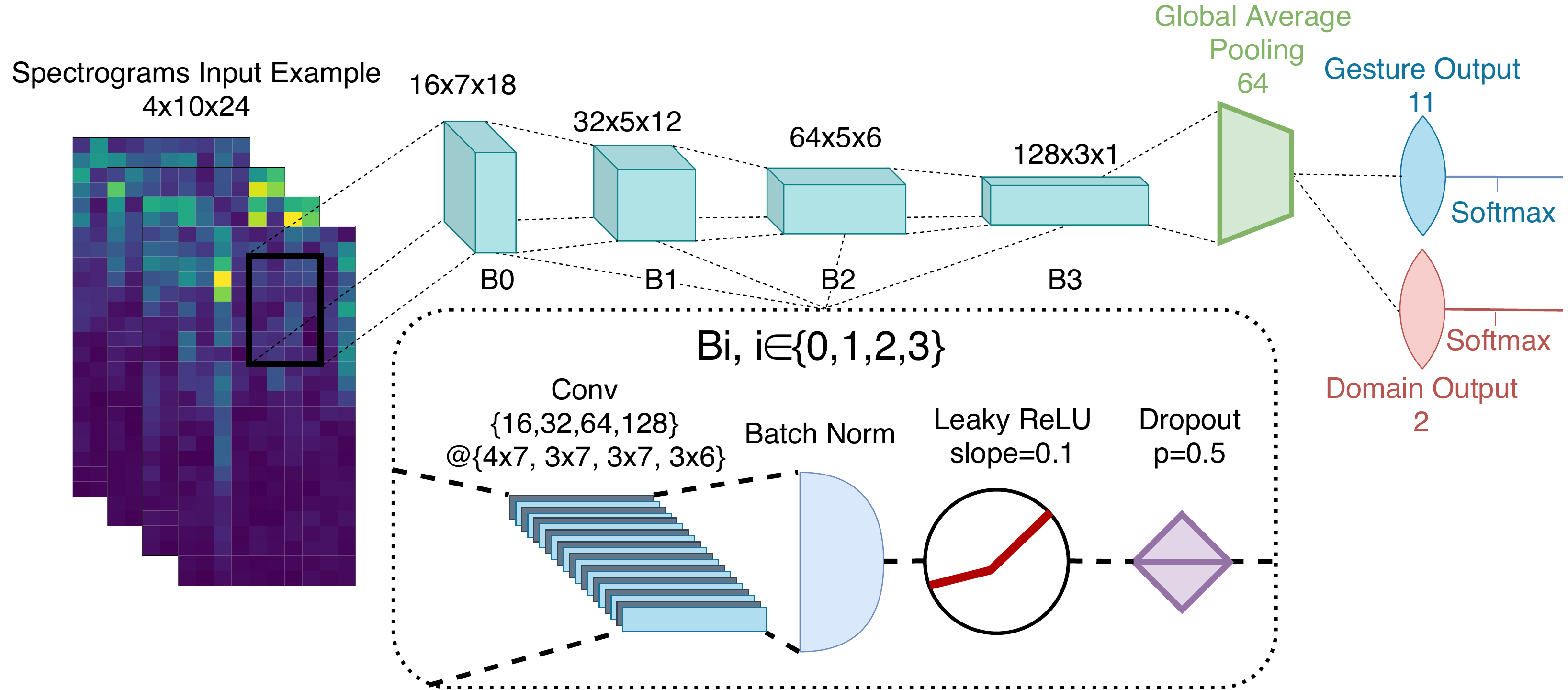}
\caption{The ConvNet's architecture employing 206 548 learnable parameters. In this figure, $B_i$ refers to the ith block ($i\in\{0,1,2,3\}$). Conv refers to a convolutional layer. When working with the reduced dataset, the number of output neurons from the gesture-head are reduced to seven.}
\label{SpectrogramConvNetArchitecture}
\end{figure*}

%Spectrograms were selected to be fed as input to the ConvNet as they were shown to be competitive with the state-of-the-art~\cite{journal_paper_TL_ulysse, self_recalibration_majority_vote}.
A small and simple ConvNet's architecture inspired from~\cite{HandCraftVsLearnedFeaturesEMG} and presented in Figure~\ref{SpectrogramConvNetArchitecture} was selected to reduce potential confounding factors. The ConvNet's architecture contains four \textit{blocks} followed by a global average pooling and two heads. The first head is used to predict the gesture held by the participant. The second head is only activated when employing domain adversarial algorithms (see Section~\ref{domain_adaptation} and~\ref{unsupervised_recalibration} for details). Each block encapsulates a convolutional layer~\cite{deep_learning}, followed by batch normalization~\cite{batch_normalization}, leaky ReLU~\cite{leaky_relu} and dropout (set to p=0.5)~\cite{dropout}. 

ADAM~\cite{adam} is employed for the ConvNet's optimization with batch size of 512. The learning rate (lr=0.001316) was selected with the 3DC Dataset by random search~\cite{random_search_bengio} using a uniform random distribution on a logarithmic scale between $10^{-5}$ and $10^1$ and 100 candidates (each candidate was evaluated 5 times). Early stopping, with a patience of 10 epochs, is also applied by using 10\% of the training dataset as a validation set. Additionally, learning rate annealing, with a factor of five and a patience of five, was also used. Within this paper, this classifier will be refered to as Spectrogram ConvNet.

%The architecture choices and hyperparameter selections were derived from the 3DC Dataset and previous literature using it (mainly~\cite{HandCraftVsLearnedFeaturesEMG, 3dcArmband}). 

Note that the ConvNet's architecture implementation, written with PyTorch~\cite{pytorch}, is made \href{https://github.com/UlysseCoteAllard/LongTermEMG}{readily available here (https://github.com/UlysseCoteAllard/LongTermEMG)}. 

\subsection{Temporal-Spatial Descriptors Deep Network}
Due to the ubiquity of handcrafted feature sets within the field of EMG-based gesture recognition, a deep network taking Temporal-Spatial Descriptors (TSD) as input is also considered. TSD is a handcrafted feature set proposed by Khushaba et al.~\cite{TSD} which achieved state-of-the-art results in EMG-based gesture classification. A short overview of this feature set is given in Appendix~\ref{appendix_comparisons_with_feature_sets} and the interested reader is encouraged to consult~\cite{TSD} for a detailed description. Note that before computing the gesture, the data is preprocessed as described in Section~\ref{preprocessing_section} (without the spectrogram part).

The deep network architecture was again selected to be as simple as possible and is comprised of 3 fully connected layers each 200 neurons wide. Each layer also applies batch normalization, leaky ReLU (slope 0.1) as the activation function and dropout (p=0.5). The training procedure is the same as for the Spectrogram ConvNet. ADAM is also employed with a learning rate of 0.002515 (found by cross-validation on the 3DC Dataset using the same hyperparameter as the Spectrogram ConvNet). The PyTorch implementation of the Deep Network, which will be referred to as TSD DNN for the remainder of this paper, is also made \href{https://github.com/UlysseCoteAllard/LongTermEMG}{readily available here (https://github.com/UlysseCoteAllard/LongTermEMG)}.

\subsection{Calibration Methods}
This work considers three types of calibration for long-term classification of sEMG signals: No Calibration, Recalibration and Unsupervised Calibration. In the first case, the network is trained solely from the data of the first session. In the Recalibration case, the model is re-trained at each new session with the new labeled data. Unsupervised Calibration is similar to Recalibration, but the dataset used for recalibration is unlabeled. Section~\ref{domain_adaptation} and~\ref{unsupervised_recalibration} presents the unsupervised calibration algorithms considered in this work. 

\section{Unsupervised Domain Adaptation}
\label{domain_adaptation}

Domain adaptation is an area in machine learning which aims at learning a discriminative predictor from two datasets (source and target datasets) coming from two different, but related, distributions~\cite{DANN} (referred to as $\mathcal{D}_s$ and $\mathcal{D}_t$). In the unsupervised case, one of the datasets is labeled (and comes from $\mathcal{D}_s$), while the second is unlabeled (and comes from $\mathcal{D}_t$). 

Within the context of myoelectric control systems, labeled data is obtained through a user's conscious calibration session. However, due to the transient nature of sEMG signals~\cite{prosthethics_review_erik, non_stationnary_EMG}, classification performance tends to degrade over time. This naturally creates a burden for the user who needs to periodically recalibrate the system to maintain its usability~\cite{non_stationnary_EMG, AdaBN}. During normal usage, however, unlabeled data is constantly generated. Consequently, the unsupervised domain adaptation setting naturally arises by defining the \textit{source dataset} as the labeled data of the calibration session and the \textit{target dataset} as the unlabeled data generated by the user during control. 

The PyTorch implementation of the domain adversarial algorithms is mainly based on~\cite{vada_dirt_T_implementation}. 

\subsection{Domain-Adversarial Training of Neural Networks}

The Domain-Adversarial Neural Network (DANN) algorithm proposes to predict on the target dataset by learning a representation from the source dataset that makes it hard to distinguish examples from either distribution~\cite{dann_OG, DANN}. To achieve this objective, DANN adds a second head (which may be comprised of one or more layers) to the network. This head, referred to as the \textit{domain classification head}, receives the features from the last feature extraction layer of the network (in this work case; from the global average pooling layer). The goal of this second head is to learn to discriminate between the two domains (source and target). However, during backpropagation, the gradient computed from the domain loss is multiplied by a negative constant (-1 in this work). This gradient reversal explicitly forces the feature distribution of the domains to be similar. The backpropagation algorithm proceeds normally for the original head (classification head). The two losses are combined as follows: $\mathcal{L}_y(\theta;\mathcal{D}_s) + \lambda_d\mathcal{L}_d(\theta; \mathcal{D}_s, \mathcal{D}_t)$, where $\theta$ is the classifier's parametrization, $\mathcal{L}_y$ and $\mathcal{L}_d$ are the prediction and domain loss respectively. $\lambda_d$ is a scalar that weights the domain loss (set to $0.1$ in this work).

\subsection{Decision-boundary Iterative Refinement Training with a Teacher}

Decision-boundary Iterative Refinement Training with a Teacher (DIRT-T) is a two-step domain-adversarial training algorithm which achieves state-of-the-art results on a variety of domain adaptation benchmarks~\cite{dirt_T}. 

\subsubsection{First step}

During the first step, referred to as VADA (for Virtual Adversarial Domain Adaptation)~\cite{dirt_T}), training is done using DANN as described previously (i.e. using a second head to discriminate between domains). However, with VADA, the network is also penalized when it violates the cluster assumption on the target. This assumption states that data belonging to the same cluster in the feature space share the same class. Consequently, decision boundaries should avoid crossing dense regions. As shown in~\cite{cluster_assumption_conditional_entropy}, this behavior can be achieved by minimizing the conditional entropy with respect to the target distribution:

\begin{equation}
    \mathcal{L}_c(\theta;\mathcal{D}_t) = \mathbb{E}_{x\sim\mathcal{D}_t}\left[ h_\theta (x)^T \ln(h_\theta(x))\right]
\end{equation}

Where $\theta$ is the parametrization of a classifier $h$.

In practice, $\mathcal{L}_c$ must be estimated from the available data. However, as noted by~\cite{cluster_assumption_conditional_entropy}, such an approximation breaks if the classifier $h$ is not locally-Lipschitz (i.e. an arbitrary small change in the classifier's input produces an arbitrarily large change in the classifier's output). To remedy this, VADA proposes to explicitly incorporate the locally-Lipschitz constraint during training via Virtual Adversarial Training (VAT)~\cite{VAT}. VAT generates new "virtual" examples at each training batch by applying small perturbation to the original data. The average maximal Kullback-Leibler divergence ($\mathbf{D_{KL}}$)~\cite{KL_divergence} is then minimized between the real and virtual examples to enforce the locally-Lipschitz constraint. In other words, VAT adds the following function to minimize during training:

\begin{equation}
    \mathcal{L}_v(\theta;\mathcal{D}) = \mathbb{E}_{x\sim\mathcal{D}}\left[ \max\limits_{||r|| \leq \epsilon}\mathbf{D_{KL}}( h_\theta (x)|| h_\theta(x+r))\right]
\end{equation}

As VAT can be seen as a form of regularization, it is also applied for the source data. In summary, the combined loss function to minimize during VADA training is:

\begin{equation}
\begin{aligned}
    \min\limits_{\theta} \mathcal{L}_y(\theta;\mathcal{D}_s) + \lambda_d\mathcal{L}_d(\theta; \mathcal{D}_s, \mathcal{D}_t) + \lambda_{vs}\mathcal{L}_v(\theta;\mathcal{D}_s) + \\
   \lambda_{vt}\mathcal{L}_v(\theta;\mathcal{D}_t) + \lambda_{c}\mathcal{L}_c(\theta;\mathcal{D}_t)
\end{aligned}
\end{equation}

\begin{figure}[!htbp]
\centering
\includegraphics[width=\linewidth]{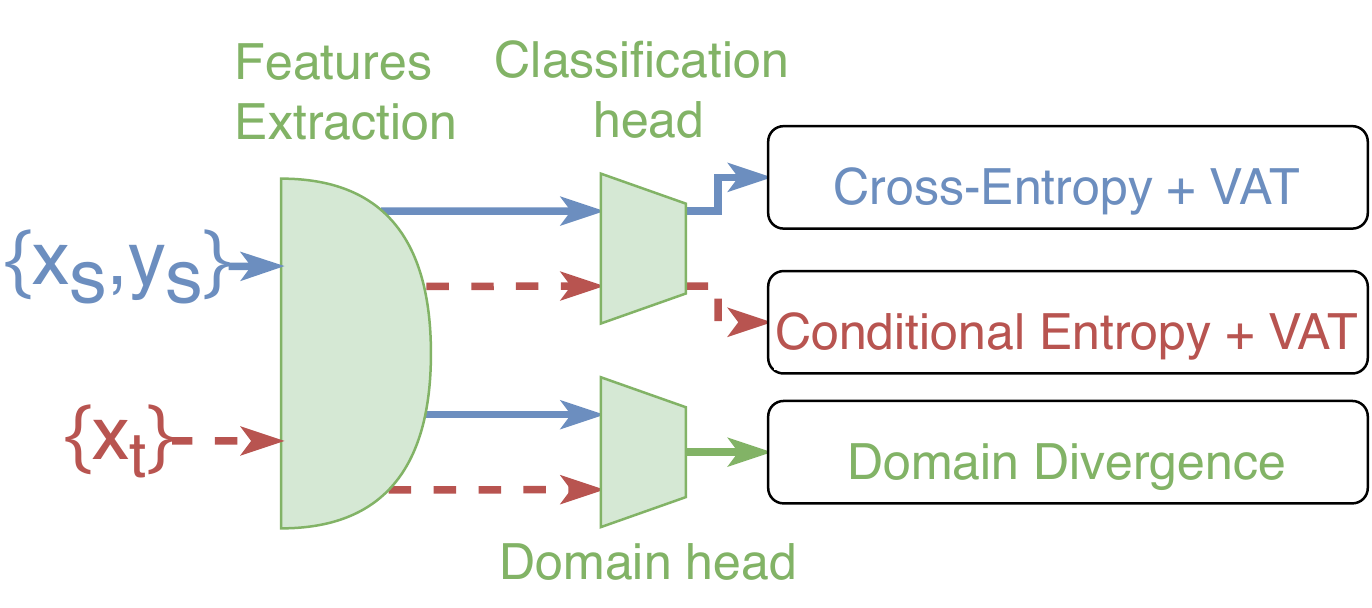}
\caption{The VADA algorithm which simultaneously tries to reduce the divergence between the labeled source (\{$x_s, y_s$\}) and unlabeled target ($\{x_t\}$) dataset while also penalizing violation of the cluster assumption on the target dataset.  }
\label{vadaDiagram}
\end{figure}

Where the importance of each additional loss function is weighted with a hyperparameter ($\lambda_d$, $\lambda_{vs}$, $\lambda_{vt}$, $\lambda_{c}$) . A diagram of VADA is provided in Figure~\ref{vadaDiagram}.

\subsubsection{Second Step}

During the second step, the signal from the source is removed. The idea is then to find a new parametrization that further minimizes the target cluster assumption violation while remaining close to the classifier found during the first step. This process can then be repeated by updating the original classifier with the classifier's parametrization found at each iteration. The combined loss function to minimize during the $n$th iteration thus becomes:

\begin{equation}
\begin{aligned}
    \min\limits_{\theta_n} \beta\mathbb{E}\left[ \mathbf{D_{KL}}( h_{\theta_{n-1}} (x)|| h_{\theta_{n}}(x))\right] +\\ \lambda_{vt}\mathcal{L}_v(\theta;\mathcal{D}_t) + \lambda_{c}\mathcal{L}_c(\theta;\mathcal{D}_t)
\end{aligned}
\label{dirt_t_equation}
\end{equation}

Where $\beta$ is a hyperparameter which weighs the importance of remaining close to $h_{\theta_{n-1}}$. In practice, the optimization problem of Eq.~\ref{dirt_t_equation} can be approximately solved with a finite number of stochastic gradient descent steps~\cite{dirt_T}. Following~\cite{dirt_T}, the hyperparameters values are set to $\lambda_d=10^{-2}$, $\lambda_{vs}=1$, $\lambda_{vt}=10^{-2}$, $\lambda_{c}=10^{-2}$, $\beta=10^{-2}$.

Note that, both DANN and VADA were conservative domain adaptation algorithms (i.e. the training algorithms try to generate a classifier that is able to discriminate between classes from both the source and target simultaneously). In contrast, DIRT-T is non-conservative as it ignores the source's signal during training. In the case where the gap between the source and the target is important, this type of non-conservative algorithm is expected to perform better than its conservative counterparts~\cite{dirt_T}. %In principle, this second step could be applied as a refinement step to any other domain adaptation training algorithms. 

\subsection{Unsupervised Adaptation - Hyperparameters Selection}
One challenge in applying unsupervised domain adaptation algorithms is the selection of the hyperparameters associated with the loss functions' weights. This is due to the absence of labeled data on the target dataset, which in practice prohibits performing standard hyperparameter selection. One possible solution is to perform the adaptation without explicitly minimizing the distance between the source and target, so that this distance can be used as a measure of adaptation performance~\cite{unsupervised_domain_adaptation_using_multitask_self_supervision}. However, such a solution precludes algorithms like the ones considered in this work and so the question of how to best perform hyperparameters selection remains a difficult and open question.

In their work introducing VADA and DIRT-T~\cite{dirt_T}, Shu et al. observed that extensive hyperparameter tuning was not necessary to achieve state-of-the-art performance on the datasets they were using. Consequently, following this observation, the hyperparameters associated with the unsupervised domain adversarial algorithms described in this section used the defaults weights recommended in their respective paper.

\section{Unsupervised Self-Calibration}
\label{unsupervised_recalibration}
Within an unsupervised domain adaptation setting, the classifier's performance is limited by the unavailability of labeled data from the target domain. However, real-time EMG-based gesture recognition offers a particular situation from which pseudo-labels can be generated from the recorded data by looking at the prediction's context. % of the classifier. 
These pseudo-labels can then be used as a way for the classifier to perform self-recalibration. Zhai et al.~\cite{self_recalibration_majority_vote} proposed to leverage this context by relabeling the network's predictions. Let $P(i, j)$ be the softmax value of the network's output for the $j$th gesture (associated with the $j$th output neuron) of the $i$th example of a sequence. The heuristic considers an array composed of the $t$ segments surrounding example $i$ (included). For each $j$, the median softmax value over this array is computed: 

\begin{equation}
\begin{aligned}
    \Tilde{P}(i, j)=median(P(i-t, j), P(i-t+1, j), ...,\\
    P(i, j), ..., P(i+t, j))
\end{aligned}
\end{equation}

The pseudo-label of $i$ then becomes the gesture $j$ associated with the maximal $\Tilde{P}(i, j)$. The median of the softmax's outputs is used instead of the prediction's mean to reduce the impact of outliers~\cite{self_recalibration_majority_vote}. This self-calibrating heuristic will be referred to as MV (for Multiple Votes) from now on. As it was the best performing setting, the All-Session recalibration setting (i.e. using all available unlabeled data across sessions)~\cite{self_recalibration_majority_vote} is employed for MV. The hyperparameter $t$ was set to 1 second, as recommended in~\cite{self_recalibration_majority_vote}. 

This work proposes to improve on MV with a new self-calibrating algorithm, named SCADANN, which can be divided into three steps:
\begin{enumerate}
    \item Apply DANN to the network using the labeled and newly acquired unlabeled data.
    \item Using the adapted network, perform the relabeling scheme described in Section~\ref{pseudo_labels}.
    \item Starting from the adapted network, train the network with the pseudo-labeled data and labeled data while continuing to apply DANN to minimize domain divergence. 
\end{enumerate}

The first step aims at reducing the domain divergence between the labeled recording session and the unlabeled recording to improve classification performance of the network. 

The second step uses the pseudo-labeling heuristic described in Section~\ref{pseudo_labels}. In addition to using the prediction's context to enhance the relabeling process, the proposed heuristic introduces two improvements compared to~\cite{self_recalibration_majority_vote}:

First, the heuristic tries to detect transition from one gesture to another. Then, already relabeled predictions falling within the transition period are vetted and possibly relabeled to better reflect when the actual transition occurred. This improvement aims at addressing two problems. First, the added latency introduced by majority-voting pseudo-labeling is removed. Second, this relabeling can provide the training algorithm with gesture transition examples. This is of particular interest as labeled transition examples are simply too time consuming to produce, especially considering the current need for periodic recalibration ($g$ gestures create $g\times (g-1)$ transitions to record).
%In fact, given a dataset with $g$ gestures, the number of transitions which would need to be recorded is $g\times (g-1)$, something that is simply not viable considering the current need for periodic recalibration. 
Introducing pseudo-labeled transition examples within the target dataset, could allow the network to detect transitions more rapidly and thus reduce the system latency. In turn, due to this latency's reduction, window's length could be increases to improve the overall system's performance.

The second improvement, introduces the notion of \textit{stability} to the network's predictions. Using this notion, the heuristic removes examples that are more likely to be relabeled falsely from the pseudo-labeled dataset. This second improvement is essential for a realistic implementation of self-calibrating algorithms, as otherwise the pseudo-labeled dataset would rapidly be filled with an important quantity of noise. This would result in a rapidly degenerating network as self-calibration is performed iteratively.

The third step re-calibrates the network using the labeled and pseudo-labeled dataset in conjunction. DANN is again employed to try to obtain a similar feature representation between the source and target datasets. The source dataset contains the labeled dataset alongside all the pseudo-labeled data from prior sessions, while the target dataset contains the pseudo-labeled data from the current session. The difference with SCADANN's first step is that the network's weights are also optimized in relation to the cross-entropy loss calculated from the newly generated pseudo-labels.
If only the pseudo-labeled dataset was employed for recalibration, the network performance would rapidly degrade from being trained only with noisy labels and possibly without certain gestures (i.e. nothing ensure that the pseudo-labeled dataset is balanced or even contains all the gestures). Early stopping is performed using part of the newly generated pseudo-labels.

\subsection{Proposed Pseudo-labels Generating Heuristic}
\label{pseudo_labels}
For concision's sake, the pseudo-code for the proposed relabeling heuristic is presented in Appendix~\ref{pseudo_labels_appendix}-Algorithm~\ref{pseudo_labeling_heuristic}. Note also that a python implementation of SCADANN (alongside the pseudo-labeling heuristic) is available in the previously mentioned online \href{https://github.com/UlysseCoteAllard/LongTermEMG}{repository}.

The main idea behind the heuristic is that if the new prediction is different than the previous one, the state goes from \textit{stable} to \textit{unstable}. 
%The main idea behind the heuristic is to look at the network's prediction one after the other, so that when the next prediction is different than the previous, the state goes from \textit{stable} to \textit{unstable}. 
%the heuristic goes from the \textit{stable state} to the \textit{unstable state}. 
During the stable state, the prediction of the considered segment is added to the pseudo-label array. During the unstable state, all the network's output (after the softmax layer) are instead accumulated in a second array. When this second array contains enough segments (hyperparameter sets to 1.5s), the class associated with the output neuron with the highest median value is defined as the new possible stable class. The new possible stable class is confirmed if the median percentage of this class (compared with the other classes) is above a certain threshold (85\% and 65\% for the seven and eleven gestures dataset respectively (selected using the 3DC dataset)). If this threshold is not achieved, the oldest element in the second array is removed and replaced with the next element. Note that the computation of the new possible stable class using the median is identical to MV. 

When the new possible class is confirmed, the heuristic first verifies if it was in the unstable state for too long (2s in this work). If it was, all the predictions accumulated during the unstable state are removed. Otherwise, if the new stable state class is different than before it means that a gesture's transition probably occurred. Consequently, the heuristic goes back in time before the instability began (maximum of 0.5s in this work) and looks at the derivative of the entropy calculated from the network's softmax output to determine when the network started to be affected by the gesture's transition. All the segments from this instability period (and adding the relevant segments from the look-back step) are then relabeled as the new stable state class found. If instead the new stable state class is identical to the previous one, only the segments from the instability period are relabeled. The heuristic then returns to its stable state.

\subsection{SCADANN - Hyperparameters Selection}
On the surface, SCADANN introduces several hyperparameters whose selection, within an unsupervised domain adaptation paradigm, is not straightforward. The majority of the introduced hyperparameters, however, have a meaningful interpretation within the context of EMG-based gesture recognition. In other words, reasonable values can be assigned to them without performing detailed data-driven hyperparameter selection. In addition, because these newly introduced hyperparameters are solely related to the pseudo-labeling aspect of the work, a labeled dataset (in this work case the 3DC Dataset) can be leveraged to perform hyperparameter selection.
\subsection{Adaptive Batch Normalization}
For the sake of completeness, in addition to the five previously mentioned adaptation algorithms, this work also considers Adaptive Batch Normalization (AdaBN)~\cite{originalAdaBN, AdaBN}. AdaBN is an unsupervised domain adaptation algorithm which was successfully applied to EMG-based gesture recognition in~\cite{AdaBN}. The hypothesis behind AdaBN is that the label-related information (the difference between gestures) can be encapsulated within the weights of the network, while the domain-related information (the difference between sessions) are contained within the batch normalization (BN) statistics. In practice, this means that the adaptation is done by feeding the unlabeled examples from the target dataset to the network to update the BN statistics. Note that within this work's setting, as only one session is contained within the source dataset and inter-user classification is not considered, the multi-stream aspect proposed in~\cite{AdaBN} cannot be applied.

\section{Experiments and results}
\label{result_section}

As suggested in~\cite{use_friedman_plus_holm}, a two-step statistical procedure is used whenever multiple algorithms are compared against each other. First, Friedman's test ranks the algorithms amongst each other. Then, Holm's post-hoc test is applied ($n=20$) using the No Calibration setting 
as a comparison basis. Additionally, Cohen's $D_z$~\cite{cohenD} is employed to determine the effect size of using one of the self-supervised algorithm over the No Calibration setting. To better contextualize the performance of the basic Spectrogram ConvNet used in this work, a comparison between the Spectrogram ConvNet and 6 widely used features ensembles within the field of sEMG-based gesture recognition is performed. For the sake of concision, this comparison is given in Appendix~\ref{appendix_comparisons_with_feature_sets}.

\subsection{Training Recording}
In this subsection, all training was performed using the first and second cycles of the relevant Training Recording, while the third cycle was employed for testing. All 20 participants completed three Training Recordings and only the labels from the first Training Recording are used (the data from the other Training Recordings are used without labels for the unsupervised recalibrations algorithms when relevant). The time-gap between each Training Recording was around seven days (14-day gap between session 1 and 3). Note that for the first session, all algorithms are equivalent to the No Calibration scheme and consequently perform the same.

\subsubsection{Offline Seven Gestures Reduced Dataset}
%The average test-set accuracy from the first Training Recording across all subjects is 93.58\%$\pm$4.58\%. This accuracy for a ConvNet is consistent with other works using spectrograms as input and the same seven gestures with similar datasets~\cite{transfer_learning_conference_emg, journal_paper_TL_ulysse}.

Table~\ref{comparison_DA_training_table_seven_gestures} shows a comparison of the No Calibration setting alongside the three DA algorithms, AdaBN, MV and SCADANN for both the Spectrogram ConvNet and the TSD DNN.

\begin{table}[ht]
\caption{Offline accuracy for seven gestures}
\label{comparison_DA_training_table_seven_gestures}
\resizebox{\linewidth}{!}{%
\begin{tabular}{cccccccc}
\hline
\multicolumn{8}{c}{\textbf{Spectrogram ConvNet}} \\ \hline
 &
  No Cal &
  DANN &
  VADA &
  Dirt-T &
  AdaBN &
  MV &
  SCADANN \\ \hline
\begin{tabular}[c]{@{}c@{}}Session 0\\ STD\end{tabular} &
  \begin{tabular}[c]{@{}c@{}}93.58\%\\ 4.58\%\end{tabular} &
  \begin{tabular}[c]{@{}c@{}}N\textbackslash{}A\\ N\textbackslash{}A\end{tabular} &
  \begin{tabular}[c]{@{}c@{}}N\textbackslash{}A\\ N\textbackslash{}A\end{tabular} &
  \begin{tabular}[c]{@{}c@{}}N\textbackslash{}A\\ N\textbackslash{}A\end{tabular} &
  \begin{tabular}[c]{@{}c@{}}N\textbackslash{}A\\ N\textbackslash{}A\end{tabular} &
  \begin{tabular}[c]{@{}c@{}}N\textbackslash{}A\\ N\textbackslash{}A\end{tabular} &
  \begin{tabular}[c]{@{}c@{}}N\textbackslash{}A\\ N\textbackslash{}A\end{tabular} \\ \hline
\begin{tabular}[c]{@{}c@{}}Session 1\\ STD\\ Friedman Rank\\ H0\\ Cohen's Dz\end{tabular} &
  \begin{tabular}[c]{@{}c@{}}71.10\%\\ 22.90\%\\ 4.85\\ N\textbackslash{}A\\ N\textbackslash{}A\end{tabular} &
  \begin{tabular}[c]{@{}c@{}}72.76\%\\ 26.00\%\\ 4.73\\ 1\\ 0.19\end{tabular} &
  \begin{tabular}[c]{@{}c@{}}73.35\%\\ 25.48\%\\ 4.25\\ 1\\ 0.24\end{tabular} &
  \begin{tabular}[c]{@{}c@{}}74.28\%\\ 24.42\%\\ 3.70\\ 1\\ 0.36\end{tabular} &
  \begin{tabular}[c]{@{}c@{}}72.61\%\\ 25.95\%\\ 5.23\\ 1\\ 0.16\end{tabular} &
  \begin{tabular}[c]{@{}c@{}}74.45\%\\ 24.03\%\\ 2.78\\ 0 (0.01193)\\ 0.62\end{tabular} &
  \textbf{\begin{tabular}[c]{@{}c@{}}75.50\%\\ 25.41\%\\ 2.48\\ 0 (0.00305)\\ 0.52\end{tabular}} \\ \hline
\begin{tabular}[c]{@{}c@{}}Session 2\\ STD\\ Friedman Rank\\ H0\\ Cohen's Dz\end{tabular} &
  \begin{tabular}[c]{@{}c@{}}68.75\%\\ 22.58\%\\ 5.60\\ N\textbackslash{}A\\ N\textbackslash{}A\end{tabular} &
  \begin{tabular}[c]{@{}c@{}}74.49\%\\ 22.73\%\\ 4.40\\ 1\\ 0.73\end{tabular} &
  \begin{tabular}[c]{@{}c@{}}75.55\%\\ 22.76\%\\ 4.03\\ 1\\ 0.77\end{tabular} &
  \begin{tabular}[c]{@{}c@{}}75.52\%\\ 23.55\%\\ 3.40\\ 0 (0.00512)\\ 0.70\end{tabular} &
  \begin{tabular}[c]{@{}c@{}}76.02\%\\ 23.10\%\\ 2.95\\ 0 (0.00063)\\ 0.79\end{tabular} &
  \begin{tabular}[c]{@{}c@{}}70.01\%\\ 24.82\%\\ 4.68\\ 1\\ 0.22\end{tabular} &
  \textbf{\begin{tabular}[c]{@{}c@{}}77.22\%\\ 22.50\%\\ 2.95\\ 0 (0.00063)\\ 0.92\end{tabular}} \\ \hline
\multicolumn{8}{c}{\textbf{TSD DNN}} \\ \hline
 &
  No Cal &
  DANN &
  VADA &
  Dirt-T &
  AdaBN &
  MV &
  SCADANN \\ \hline
\begin{tabular}[c]{@{}c@{}}Session 0\\ STD\end{tabular} &
  \begin{tabular}[c]{@{}c@{}}96.39\%\\ 3.20\%\end{tabular} &
  \begin{tabular}[c]{@{}c@{}}N\textbackslash{}A\\ N\textbackslash{}A\end{tabular} &
  \begin{tabular}[c]{@{}c@{}}N\textbackslash{}A\\ N\textbackslash{}A\end{tabular} &
  \begin{tabular}[c]{@{}c@{}}N\textbackslash{}A\\ N\textbackslash{}A\end{tabular} &
  \begin{tabular}[c]{@{}c@{}}N\textbackslash{}A\\ N\textbackslash{}A\end{tabular} &
  \begin{tabular}[c]{@{}c@{}}N\textbackslash{}A\\ N\textbackslash{}A\end{tabular} &
  \begin{tabular}[c]{@{}c@{}}N\textbackslash{}A\\ N\textbackslash{}A\end{tabular} \\ \hline
\begin{tabular}[c]{@{}c@{}}Session 1\\ STD\\ Friedman Rank\\ H0\\ Cohen's Dz\end{tabular} &
  \begin{tabular}[c]{@{}c@{}}78.14\%\\ 18.49\%\\ 5.45\\ N\textbackslash{}A\\ N\textbackslash{}A\end{tabular} &
  \begin{tabular}[c]{@{}c@{}}83.15\%\\ 15.47\%\\ 4.03\\ 1\\ 0.90\end{tabular} &
  \begin{tabular}[c]{@{}c@{}}80.90\%\\ 15.46\%\\ 4.95\\ 1\\ 0.37\end{tabular} &
  \begin{tabular}[c]{@{}c@{}}80.94\%\\ 14.06\%\\ 4.68\\ 1\\ 0.22\end{tabular} &
  \begin{tabular}[c]{@{}c@{}}84.37\%\\ 14.64\%\\ 3.00\\ 0 (0.00168)\\ 0.88\end{tabular} &
  \begin{tabular}[c]{@{}c@{}}83.01\%\\ 19.43\%\\ 3.48\\ 0 (0.01536)\\ 0.86\end{tabular} &
  \textbf{\begin{tabular}[c]{@{}c@{}}84.91\%\\ 16.09\%\\ 2.43\\ 0 (0.00006)\\ 0.84\end{tabular}} \\ \hline
\begin{tabular}[c]{@{}c@{}}Session 2\\ STD\\ Friedman Rank\\ H0\\ Cohen's Dz\end{tabular} &
  \begin{tabular}[c]{@{}c@{}}79.78\%\\ 19.06\%\\ 5.20\\ N\textbackslash{}A\\ N\textbackslash{}A\end{tabular} &
  \begin{tabular}[c]{@{}c@{}}84.73\%\\ 19.38\%\\ 3.93\\ 1\\ 0.55\end{tabular} &
  \begin{tabular}[c]{@{}c@{}}84.50\%\\ 17.37\%\\ 4.20\\ 1\\ 0.52\end{tabular} &
  \begin{tabular}[c]{@{}c@{}}82.16\%\\ 17.68\%\\ 5.18\\ 1\\ 0.28\end{tabular} &
  \begin{tabular}[c]{@{}c@{}}85.91\%\\ 19.06\%\\ 3.23\\ 0 (0.01919)\\ 0.61\end{tabular} &
  \begin{tabular}[c]{@{}c@{}}81.47\%\\ 19.23\%\\ 4.15\\ 1\\ 0.48\end{tabular} &
  \textbf{\begin{tabular}[c]{@{}c@{}}88.20\%\\ 17.55\%\\ 2.13\\ 0 (0.00004)\\ 0.81\end{tabular}} \\ \hline
\end{tabular}
}
\end{table}

\subsubsection{Offline Eleven Gesture Dataset}
\label{results_offline_eleven_gestures}
%The average test-set accuracy from the first Training Recording across all subjects is 84.19\%$\pm$9.12\%, which is consistent with accuracies obtained on the 3DC datasets~\cite{3dcArmband,HandCraftVsLearnedFeaturesEMG}.

\begin{figure}[!htbp]
\includegraphics[ width=\linewidth]{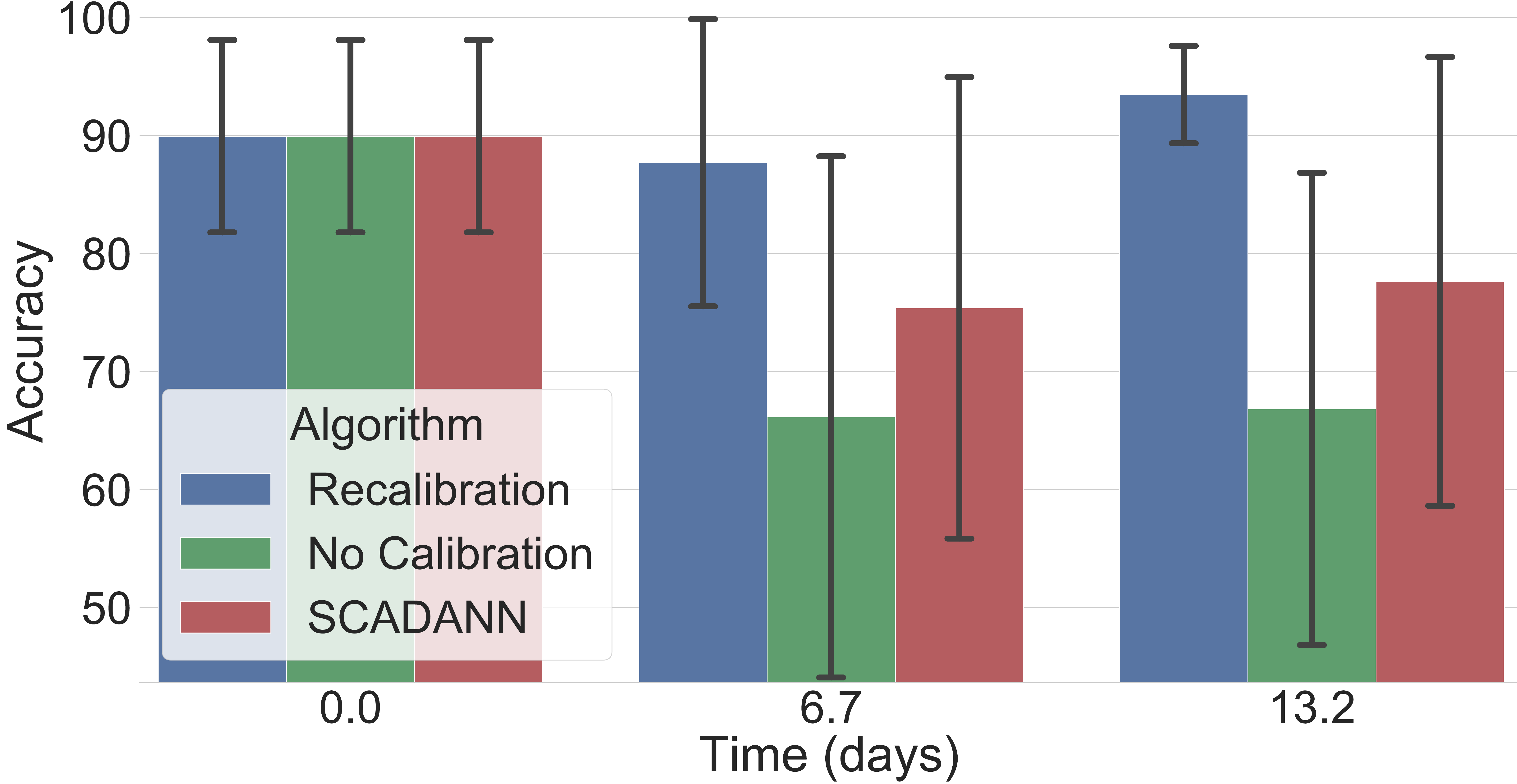}
\caption{Offline accuracy using the TSD DNN for the eleven gestures in respect to time. The values on the x-axis represent the average number of days elapsed across participants since the first session.}
\label{eleven_gestures_TSD_DNN_long_term_classification}
\end{figure}

Table~\ref{comparison_DA_training_table_Eleven_Gestures} compares the No Calibration setting with the three DA algorithms, AdaBN, MV and SCADANN for both networks.  Figure~\ref{eleven_gestures_TSD_DNN_long_term_classification} shows a histogram of the accuracy obtained by the TSD DNN for the No Calibration, SCADANN and the Recalibration methods. 

\begin{table}[ht]
\caption{Offline accuracy for eleven gestures}
\label{comparison_DA_training_table_Eleven_Gestures}
\resizebox{\linewidth}{!}{%
\begin{tabular}{cccccccc}
\hline
\multicolumn{8}{c}{\textbf{Spectrogram ConvNet}} \\ \hline
 &
  No Cal &
  DANN &
  VADA &
  Dirt-T &
  AdaBN &
  MV &
  SCADANN \\ \hline
\begin{tabular}[c]{@{}c@{}}Session 0\\ STD\end{tabular} &
  \begin{tabular}[c]{@{}c@{}}84.19\%\\ 9.12\%\end{tabular} &
  \begin{tabular}[c]{@{}c@{}}N\textbackslash{}A\\ N\textbackslash{}A\end{tabular} &
  \begin{tabular}[c]{@{}c@{}}N\textbackslash{}A\\ N\textbackslash{}A\end{tabular} &
  \begin{tabular}[c]{@{}c@{}}N\textbackslash{}A\\ N\textbackslash{}A\end{tabular} &
  \begin{tabular}[c]{@{}c@{}}N\textbackslash{}A\\ N\textbackslash{}A\end{tabular} &
  \begin{tabular}[c]{@{}c@{}}N\textbackslash{}A\\ N\textbackslash{}A\end{tabular} &
  \begin{tabular}[c]{@{}c@{}}N\textbackslash{}A\\ N\textbackslash{}A\end{tabular} \\ \hline
\begin{tabular}[c]{@{}c@{}}Session 1\\ STD\\ Friedman Rank\\ H0\\ Cohen's Dz\end{tabular} &
  \begin{tabular}[c]{@{}c@{}}58.29\%\\ 25.33\%\\ 5.50\\ N\textbackslash{}A\\ N\textbackslash{}A\end{tabular} &
  \begin{tabular}[c]{@{}c@{}}62.27\%\\ 24.86\%\\ 3.85\\ 0 (0.04626)\\ 0.63\end{tabular} &
  \begin{tabular}[c]{@{}c@{}}62.45\%\\ 25.00\%\\ 3.83\\ 0 (0.04626)\\ 0.63\end{tabular} &
  \begin{tabular}[c]{@{}c@{}}62.35\%\\ 24.99\%\\ 3.78\\ 0 (0.04626)\\ 0.57\end{tabular} &
  \begin{tabular}[c]{@{}c@{}}61.83\%\\ 25.42\%\\ 4.05\\ 0 (0.04626)\\ 0.49\end{tabular} &
  \begin{tabular}[c]{@{}c@{}}60.75\%\\ 26.38\%\\ 3.55\\ 0 (0.02155)\\ 0.93\end{tabular} &
  \textbf{\begin{tabular}[c]{@{}c@{}}63.00\%\\ 24.84\%\\ 3.45\\ 0 (0.01615)\\ 0.71\end{tabular}} \\ \hline
\begin{tabular}[c]{@{}c@{}}Session 2\\ STD\\ Friedman Rank\\ H0\\ Cohen's Dz\end{tabular} &
  \begin{tabular}[c]{@{}c@{}}56.69\%\\ 23.04\%\\ 5.43\\ N\textbackslash{}A\\ N\textbackslash{}A\end{tabular} &
  \begin{tabular}[c]{@{}c@{}}62.08\%\\ 22.84\%\\ 3.95\\ 1\\ 0.75\end{tabular} &
  \begin{tabular}[c]{@{}c@{}}62.40\%\\ 22.77\%\\ 3.65\\ 0 (0.04684)\\ 0.77\end{tabular} &
  \begin{tabular}[c]{@{}c@{}}62.43\%\\ 22.69\%\\ 3.68\\ 0 (0.04684)\\ 0.75\end{tabular} &
  \begin{tabular}[c]{@{}c@{}}62.49\%\\ 22.98\%\\ 3.80\\ 1\\ 0.78\end{tabular} &
  \begin{tabular}[c]{@{}c@{}}58.27\%\\ 23.26\%\\ 4.45\\ 1\\ 0.53\end{tabular} &
  \textbf{\begin{tabular}[c]{@{}c@{}}63.43\%\\ 23.03\%\\ 3.05\\ 0 (0.00305)\\ 0.68\end{tabular}} \\ \hline
\multicolumn{8}{c}{\textbf{TSD DNN}} \\ \hline
 &
  No Cal &
  DANN &
  VADA &
  Dirt-T &
  AdaBN &
  MV &
  SCADANN \\ \hline
\begin{tabular}[c]{@{}c@{}}Session 0\\ STD\end{tabular} &
  \begin{tabular}[c]{@{}c@{}}89.95\%\\ 8.37\%\end{tabular} &
  \begin{tabular}[c]{@{}c@{}}N\textbackslash{}A\\ N\textbackslash{}A\end{tabular} &
  \begin{tabular}[c]{@{}c@{}}N\textbackslash{}A\\ N\textbackslash{}A\end{tabular} &
  \begin{tabular}[c]{@{}c@{}}N\textbackslash{}A\\ N\textbackslash{}A\end{tabular} &
  \begin{tabular}[c]{@{}c@{}}N\textbackslash{}A\\ N\textbackslash{}A\end{tabular} &
  \begin{tabular}[c]{@{}c@{}}N\textbackslash{}A\\ N\textbackslash{}A\end{tabular} &
  \begin{tabular}[c]{@{}c@{}}N\textbackslash{}A\\ N\textbackslash{}A\end{tabular} \\ \hline
\begin{tabular}[c]{@{}c@{}}Session 1\\ STD\\ Friedman Rank\\ H0\\ Cohen's Dz\end{tabular} &
  \begin{tabular}[c]{@{}c@{}}66.16\%\\ 22.66\%\\ 5.65\\ N\textbackslash{}A\\ N\textbackslash{}A\end{tabular} &
  \begin{tabular}[c]{@{}c@{}}72.44\%\\ 20.58\%\\ 3.83\\ 0 (0.02265)\\ 0.76\end{tabular} &
  \begin{tabular}[c]{@{}c@{}}69.25\%\\ 19.51\%\\ 4.75\\ 1\\ 0.36\end{tabular} &
  \begin{tabular}[c]{@{}c@{}}69.14\%\\ 16.64\%\\ 4.88\\ 1\\ 0.26\end{tabular} &
  \begin{tabular}[c]{@{}c@{}}73.63\%\\ 19.79\%\\ 3.18\\ 0 (0.00146)\\ 0.87\end{tabular} &
  \begin{tabular}[c]{@{}c@{}}71.34\%\\ 23.41\%\\ 3.33\\ 0 (0.00266)\\ 0.92\end{tabular} &
  \textbf{\begin{tabular}[c]{@{}c@{}}75.40\%\\ 20.06\%\\ 2.40\\ 0 (0.00001)\\ 1.10\end{tabular}} \\ \hline
\begin{tabular}[c]{@{}c@{}}Session 2\\ STD\\ Friedman Rank\\ H0\\ Cohen's Dz\end{tabular} &
  \begin{tabular}[c]{@{}c@{}}66.84\%\\ 20.53\%\\ 6.15\\ N\textbackslash{}A\\ N\textbackslash{}A\end{tabular} &
  \begin{tabular}[c]{@{}c@{}}74.30\%\\ 20.57\%\\ 4.13\\ 0 (0.00607)\\ 0.82\end{tabular} &
  \begin{tabular}[c]{@{}c@{}}73.61\%\\ 18.65\%\\ 3.75\\ 0 (0.00177)\\ 0.71\end{tabular} &
  \begin{tabular}[c]{@{}c@{}}73.71\%\\ 17.26\%\\ 3.95\\ 0 (0.00384)\\ 0.63\end{tabular} &
  \begin{tabular}[c]{@{}c@{}}74.99\%\\ 21.97\%\\ 2.98\\ 0 (0.00002)\\ 0.80\end{tabular} &
  \begin{tabular}[c]{@{}c@{}}69.94\%\\ 20.19\%\\ 4.70\\ 0 (0.03379)\\ 1.02\end{tabular} &
  \textbf{\begin{tabular}[c]{@{}c@{}}77.65\%\\ 19.52\%\\ 2.35\\ 0 (\textless{}0.00001)\\ 1.12\end{tabular}} \\ \hline
\end{tabular}}
\end{table}

\subsection{Evaluation Recording}
%In this subsection, training using labeled data were conducted using the first, second and third cycles of the relevant Training Recording. Adaptation was performed using the first Evaluation Recording of each session, while testing was done using the corresponding second Evaluation Recording of each session.
\subsubsection{Eleven Gestures - Dynamic Dataset, offline adaptation}
Table~\ref{comparison_DA_training_offline_test_online} compares the No Calibration setting with the three DA algorithms, AdaBN, MV and SCADANN for both networks on the second Evaluation Recording of each session, when the labeled and unlabeled data leveraged for training comes from the Training Recordings (as in Section~\ref{results_offline_eleven_gestures}).

\begin{table}[ht]
\caption{Dynamic dataset's accuracy for eleven gestures using Training Recordings as unlabeled data}
\label{comparison_DA_training_offline_test_online}
\resizebox{\linewidth}{!}{%
\begin{tabular}{cccccccc}
\hline
\multicolumn{8}{c}{\textbf{Spectrogram ConvNet}} \\ \hline
 &
  No Cal &
  DANN &
  VADA &
  Dirt-T &
  AdaBN &
  MV &
  SCADANN \\ \hline
\begin{tabular}[c]{@{}c@{}}Session 0\\ STD\end{tabular} &
  \begin{tabular}[c]{@{}c@{}}47.81\%\\ 10.94\%\end{tabular} &
  \begin{tabular}[c]{@{}c@{}}N\textbackslash{}A\\ N\textbackslash{}A\end{tabular} &
  \begin{tabular}[c]{@{}c@{}}N\textbackslash{}A\\ N\textbackslash{}A\end{tabular} &
  \begin{tabular}[c]{@{}c@{}}N\textbackslash{}A\\ N\textbackslash{}A\end{tabular} &
  \begin{tabular}[c]{@{}c@{}}N\textbackslash{}A\\ N\textbackslash{}A\end{tabular} &
  \begin{tabular}[c]{@{}c@{}}N\textbackslash{}A\\ N\textbackslash{}A\end{tabular} &
  \begin{tabular}[c]{@{}c@{}}N\textbackslash{}A\\ N\textbackslash{}A\end{tabular} \\ \hline
\begin{tabular}[c]{@{}c@{}}Session 1\\ STD\\ Friedman Rank\\ H0\\ Cohen's Dz\end{tabular} &
  \begin{tabular}[c]{@{}c@{}}38.39\%\\ 16.65\%\\ 4.80\\ N\textbackslash{}A\\ N\textbackslash{}A\end{tabular} &
  \begin{tabular}[c]{@{}c@{}}39.64\%\\ 17.37\%\\ 3.78\\ 1\\ 0.45\end{tabular} &
  \begin{tabular}[c]{@{}c@{}}39.52\%\\ 17.66\%\\ 4.10\\ 1\\ 0.32\end{tabular} &
  \begin{tabular}[c]{@{}c@{}}39.07\%\\ 17.56\%\\ 4.70\\ 1\\ 0.16\end{tabular} &
  \begin{tabular}[c]{@{}c@{}}38.99\\ 17.16\%\\ 4.33\\ 1\\ 0.17\end{tabular} &
  \begin{tabular}[c]{@{}c@{}}39.70\%\\ 17.75\%\\ 3.30\\ 1\\ 0.54\end{tabular} &
  \textbf{\begin{tabular}[c]{@{}c@{}}40.80\%\\ 17.77\%\\ 3.00\\ 1\\ 0.63\end{tabular}} \\ \hline
\begin{tabular}[c]{@{}c@{}}Session 2\\ STD\\ Friedman Rank\\ H0\\ Cohen's Dz\end{tabular} &
  \begin{tabular}[c]{@{}c@{}}38.54\%\\ 14.65\%\\ 5.50\\ N\textbackslash{}A\\ N\textbackslash{}A\end{tabular} &
  \begin{tabular}[c]{@{}c@{}}39.87\%\\ 15.32\%\\ 4.20\\ 1\\ 0.33\end{tabular} &
  \begin{tabular}[c]{@{}c@{}}40.07\%\\ 15.81\%\\ 3.60\\ 0 (0.02166)\\ 0.35\end{tabular} &
  \begin{tabular}[c]{@{}c@{}}39.59\%\\ 15.43\%\\ 4.45\\ 1\\ 0.25\end{tabular} &
  \begin{tabular}[c]{@{}c@{}}39.53\%\\ 15.59\%\\ 4.30\\ 1\\ 0.22\end{tabular} &
  \begin{tabular}[c]{@{}c@{}}40.98\%\\ 15.18\%\\ 3.35\\ 0 (0.00824)\\ 0.84\end{tabular} &
  \textbf{\begin{tabular}[c]{@{}c@{}}42.26\%\\ 16.34\%\\ 2.60\\ 0 (0.00013)\\ 0.65\end{tabular}} \\ \hline
\multicolumn{8}{c}{\textbf{TSD DNN}} \\ \hline
 &
  No Cal &
  DANN &
  VADA &
  Dirt-T &
  AdaBN &
  MV &
  SCADANN \\ \hline
\begin{tabular}[c]{@{}c@{}}Session 0\\ STD\end{tabular} &
  \begin{tabular}[c]{@{}c@{}}53.08\%\\ 11.48\%\end{tabular} &
  \begin{tabular}[c]{@{}c@{}}N\textbackslash{}A\\ N\textbackslash{}A\end{tabular} &
  \begin{tabular}[c]{@{}c@{}}N\textbackslash{}A\\ N\textbackslash{}A\end{tabular} &
  \begin{tabular}[c]{@{}c@{}}N\textbackslash{}A\\ N\textbackslash{}A\end{tabular} &
  \begin{tabular}[c]{@{}c@{}}N\textbackslash{}A\\ N\textbackslash{}A\end{tabular} &
  \begin{tabular}[c]{@{}c@{}}N\textbackslash{}A\\ N\textbackslash{}A\end{tabular} &
  \begin{tabular}[c]{@{}c@{}}N\textbackslash{}A\\ N\textbackslash{}A\end{tabular} \\ \hline
\begin{tabular}[c]{@{}c@{}}Session 1\\ STD\\ Friedman Rank\\ H0\\ Cohen's Dz\end{tabular} &
  \begin{tabular}[c]{@{}c@{}}46.09\%\\ 14.70\%\\ 4.50\\ N\textbackslash{}A\\ N\textbackslash{}A\end{tabular} &
  \begin{tabular}[c]{@{}c@{}}48.07\%\\ 14.59\%\\ 3.00\\ 1\\ 0.55\end{tabular} &
  \begin{tabular}[c]{@{}c@{}}43.92\%\\ 13.45\%\\ 5.40\\ 1\\ -0.37\end{tabular} &
  \begin{tabular}[c]{@{}c@{}}42.75\%\\ 12.65\%\\ 5.80\\ 1\\ -0.49\end{tabular} &
  \begin{tabular}[c]{@{}c@{}}47.11\%\\ 14.11\%\\ 3.75\\ 1\\ 0.21\end{tabular} &
  \begin{tabular}[c]{@{}c@{}}48.36\%\\ 14.30\%\\ 2.85\\ 1\\ 1.08\end{tabular} &
  \textbf{\begin{tabular}[c]{@{}c@{}}49.09\%\\ 14.68\%\\ 2.70\\ 1\\ 0.67\end{tabular}} \\ \hline
\begin{tabular}[c]{@{}c@{}}Session 2\\ STD\\ Friedman Rank\\ H0\\ Cohen's Dz\end{tabular} &
  \begin{tabular}[c]{@{}c@{}}46.01\%\\ 15.72\%\\ 4.90\\ N\textbackslash{}A\\ N\textbackslash{}A\end{tabular} &
  \begin{tabular}[c]{@{}c@{}}48.50\%\\ 15.80\%\\ 3.85\\ 1\\ 0.50\end{tabular} &
  \begin{tabular}[c]{@{}c@{}}45.69\%\\ 14.21\%\\ 4.80\\ 1\\ -0.05\end{tabular} &
  \begin{tabular}[c]{@{}c@{}}45.48\%\\ 13.26\%\\ 5.00\\ 1\\ -0.08\end{tabular} &
  \begin{tabular}[c]{@{}c@{}}48.35\%\\ 16.17\%\\ 3.73\\ 1\\ 0.42\end{tabular} &
  \begin{tabular}[c]{@{}c@{}}48.17\%\\ 17.06\%\\ 3.78\\ 1\\ 0.60\end{tabular} &
  \textbf{\begin{tabular}[c]{@{}c@{}}50.90\%\\ 16.64\%\\ 1.95\\ 0 (0.00009)\\ 0.91\end{tabular}} \\ \hline
\end{tabular}}
\end{table}

%The average accuracy obtained from the Spectrogram ConvNet on the second Evaluation Recording of each experiment's session across all participants is 41.58\%$\pm$14.72\% and 49.84\%$\pm$10.93\% for the No Calibration and Recalibration setting respectively. For the TSD DNN, the accuracies increase to 48.39\%$\pm$14.24\% and 54.59\%$\pm$11.49\% for the No Calibration and Recalibration respectively.  

\subsubsection{Eleven Gestures - Adaptation on the Dynamic Dataset}

Table~\ref{comparison_DA_training_online_test_online} presents the comparison between the No Calibration setting and using the first Evaluation Recording of each experiment's session as the unlabeled dataset for the three DA algorithms, AdaBN, MV and SCADANN.

\begin{table}[ht]
\caption{Dynamic dataset's accuracy for eleven gestures using the first Evaluation Recording as unlabeled data}
\label{comparison_DA_training_online_test_online}
\resizebox{\linewidth}{!}{%
\begin{tabular}{@{}cccccccc@{}}
\toprule
\multicolumn{8}{c}{\textbf{Spectrogram ConvNet}} \\ \midrule
 &
  No Cal &
  DANN &
  VADA &
  Dirt-T &
  AdaBN &
  MV &
  SCADANN \\ \midrule
\begin{tabular}[c]{@{}c@{}}Session 0\\ STD\\ Friedman Rank\\ H0\\ Cohen's Dz\end{tabular} &
  \begin{tabular}[c]{@{}c@{}}47.81\%\\ 10.94\%\\ 4.75\\ N\textbackslash{}A\\ N\textbackslash{}A\end{tabular} &
  \begin{tabular}[c]{@{}c@{}}49.37\%\\ 11.24\%\\ 3.80\\ 1\\ 0.64\end{tabular} &
  \begin{tabular}[c]{@{}c@{}}49.36\%\\ 11.04\%\\ 3.78\\ 1\\ 0.60\end{tabular} &
  \begin{tabular}[c]{@{}c@{}}49.48\%\\ 11.21\%\\ 3.38\\ 1\\ 0.53\end{tabular} &
  \begin{tabular}[c]{@{}c@{}}47.33\%\\ 10.45\%\\ 4.95\\ 1\\ -0.11\end{tabular} &
  \begin{tabular}[c]{@{}c@{}}47.68\%\\ 11.27\%\\ 4.80\\ 1\\ -0.07\end{tabular} &
  \textbf{\begin{tabular}[c]{@{}c@{}}49.89\%\\ 11.25\%\\ 2.55\\ 0 (0.00490)\\ 0.73\end{tabular}} \\ \midrule
\begin{tabular}[c]{@{}c@{}}Session 1\\ STD\\ Friedman Rank\\ H0\\ Cohen's Dz\end{tabular} &
  \begin{tabular}[c]{@{}c@{}}38.39\%\\ 16.65\%\\ 5.15\\ N\textbackslash{}A\\ N\textbackslash{}A\end{tabular} &
  \begin{tabular}[c]{@{}c@{}}40.92\%\\ 18.51\%\\ 3.10\\ 0 (0.02643)\\ 0.56\end{tabular} &
  \begin{tabular}[c]{@{}c@{}}40.73\%\\ 18.55\%\\ 3.63\\ 1\\ 0.53\end{tabular} &
  \begin{tabular}[c]{@{}c@{}}40.66\%\\ 18.38\%\\ 3.85\\ 1\\ 0.49\end{tabular} &
  \begin{tabular}[c]{@{}c@{}}40.36\%\\ 17.77\%\\ 4.25\\ 1\\ 0.41\end{tabular} &
  \begin{tabular}[c]{@{}c@{}}38.60\%\\ 17.13\%\\ 4.83\\ 1\\ 0.13\end{tabular} &
  \textbf{\begin{tabular}[c]{@{}c@{}}41.07\%\\ 19.11\%\\ 3.20\\ 0 (0.02643)\\ 0.52\end{tabular}} \\ \midrule
\begin{tabular}[c]{@{}c@{}}Session 2\\ STD\\ Friedman Rank\\ H0\\ Cohen's Dz\end{tabular} &
  \begin{tabular}[c]{@{}c@{}}38.54\%\\ 14.65\%\\ 5.10\\ N\textbackslash{}A\\ N\textbackslash{}A\end{tabular} &
  \begin{tabular}[c]{@{}c@{}}40.78\%\\ 16.05\%\\ 2.78\\ 0 (0.00063)\\ 0.50\end{tabular} &
  \begin{tabular}[c]{@{}c@{}}40.82\%\\ 16.05\%\\ 3.28\\ 0 (0.00266)\\ 0.48\end{tabular} &
  \begin{tabular}[c]{@{}c@{}}41.01\%\\ 16.29\%\\ 2.95\\ 0 (0.00063)\\ 0.51\end{tabular} &
  \begin{tabular}[c]{@{}c@{}}38.15\%\\ 15.36\%\\ 5.60\\ 1\\ -0.07\end{tabular} &
  \begin{tabular}[c]{@{}c@{}}40.02\%\\ 15.42\%\\ 4.10\\ 1\\ 0.79\end{tabular} &
  \textbf{\begin{tabular}[c]{@{}c@{}}41.41\%\\ 16.45\%\\ 3.50\\ 0 (0.00633)\\ 0.48\end{tabular}} \\ \midrule
\multicolumn{8}{c}{\textbf{TSD DNN}} \\ \midrule
 &
  No Cal &
  DANN &
  VADA &
  Dirt-T &
  AdaBN &
  MV &
  SCADANN \\ \midrule
\begin{tabular}[c]{@{}c@{}}Session 0\\ STD\\ Friedman Rank\\ H0\\ Cohen's Dz\end{tabular} &
  \begin{tabular}[c]{@{}c@{}}53.08\%\\ 11.48\%\\ 4.10\\ N\textbackslash{}A\\ N\textbackslash{}A\end{tabular} &
  \begin{tabular}[c]{@{}c@{}}55.29\%\\ 12.09\%\\ 2.60\\ 1\\ 0.71\end{tabular} &
  \begin{tabular}[c]{@{}c@{}}50.42\%\\ 10.67\%\\ 5.80\\ 1\\ -0.81\end{tabular} &
  \begin{tabular}[c]{@{}c@{}}53.59\%\\ 11.51\%\\ 4.00\\ 1\\ 0.13\end{tabular} &
  \begin{tabular}[c]{@{}c@{}}49.98\%\\ 10.90\%\\ 5.50\\ 1\\ -0.60\end{tabular} &
  \begin{tabular}[c]{@{}c@{}}53.67\%\\ 11.51\%\\ 3.70\\ 1\\ 0.30\end{tabular} &
  \textbf{\begin{tabular}[c]{@{}c@{}}55.69\%\\ 12.37\%\\ 2.30\\ 1\\ 0.67\end{tabular}} \\ \midrule
\begin{tabular}[c]{@{}c@{}}Session 1\\ STD\\ Friedman Rank\\ H0\\ Cohen's Dz\end{tabular} &
  \begin{tabular}[c]{@{}c@{}}46.09\%\\ 14.70\%\\ 5.50\\ N\textbackslash{}A\\ N\textbackslash{}A\end{tabular} &
  \begin{tabular}[c]{@{}c@{}}50.65\%\\ 14.55\%\\ 2.30\\ 0 (0.00001)\\ 1.35\end{tabular} &
  \begin{tabular}[c]{@{}c@{}}46.10\%\\ 13.66\%\\ 5.55\\ 1\\ \textless{}0.01\end{tabular} &
  \begin{tabular}[c]{@{}c@{}}49.30\%\\ 13.81\%\\ 3.50\\ 0 (0.01366)\\ 0.73\end{tabular} &
  \begin{tabular}[c]{@{}c@{}}48.12\%\\ 14.14\%\\ 4.60\\ 1\\ 0.36\end{tabular} &
  \begin{tabular}[c]{@{}c@{}}47.34\%\\ 16.16\%\\ 4.40\\ 1\\ 0.46\end{tabular} &
  \textbf{\begin{tabular}[c]{@{}c@{}}51.41\%\\ 15.46\%\\ 2.15\\ 0 (0.00001)\\ 1.17\end{tabular}} \\ \midrule
\begin{tabular}[c]{@{}c@{}}Session 2\\ STD\\ Friedman Rank\\ H0\\ Cohen's Dz\end{tabular} &
  \begin{tabular}[c]{@{}c@{}}46.01\%\\ 15.72\%\\ 5.40\\ N\textbackslash{}A\\ N\textbackslash{}A\end{tabular} &
  \begin{tabular}[c]{@{}c@{}}50.91\%\\ 15.88\%\\ 2.65\\ 0 (0.00028)\\ 0.98\end{tabular} &
  \begin{tabular}[c]{@{}c@{}}48.33\%\\ 14.12\%\\ 4.20\\ 1\\ 0.46\end{tabular} &
  \begin{tabular}[c]{@{}c@{}}50.27\%\\ 14.60\%\\ 3.15\\ 0 (0.00396)\\ 0.84\end{tabular} &
  \begin{tabular}[c]{@{}c@{}}44.22\%\\ 14.58\%\\ 5.90\\ 1\\ -0.28\end{tabular} &
  \begin{tabular}[c]{@{}c@{}}46.90\%\\ 16.31\%\\ 4.50\\ 1\\ 0.37\end{tabular} &
  \textbf{\begin{tabular}[c]{@{}c@{}}52.01\%\\ 17.17\%\\ 2.20\\ 0 (0.00002)\\ 1.32\end{tabular}} \\ \bottomrule
\end{tabular}
}
\end{table}

\begin{figure}[!htbp]
\includegraphics[width=\linewidth]{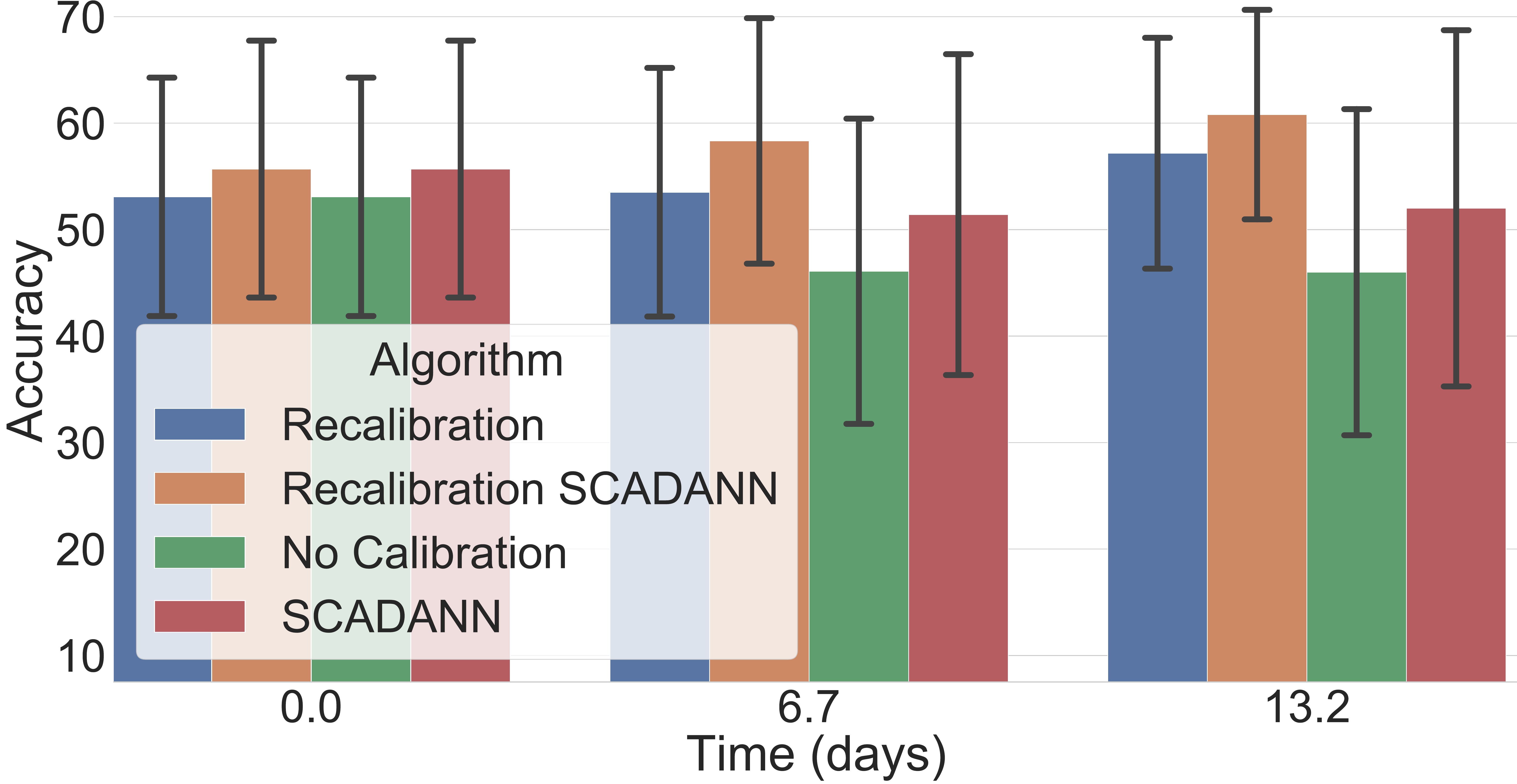}
\caption{TSD DNN dynamic dataset's accuracy for eleven gestures in respect to time. Training is performed offline with the first Training Recording session. Adaptation takes place on the first Evaluation Recording of the corresponding tested session, while the test set comes from the second Evaluation Recording of the same tested session. The values on the x-axis represent the average number of days elapsed across participants since the first session.}
\label{train_online_test_online_figure}
\end{figure}

 A histogram of the dynamic dataset's accuracy of the No Calibration, Recalibrated, SCADANN and Recalibrated SCADANN methods, trained on the TSD DNN, using the first Evaluation Recording of each experimental session as unlabeled data is shown in Figure~\ref{train_online_test_online_figure}. The Recalibration SCADANN scheme systematically and significantly (p$<$0.05) outperforms the Recalibration scheme for all three sessions for both networks, using the Wilcoxon signed rank-test~\cite{wilcoxon_signed_rank_test, use_friedman_plus_holm}, as can be seen from Table~\ref{comparisonRecalibrationAndRecal_SCADANN}.
 
\begin{table}[!htbp]
\caption{Accuracy for the \textit{Recalibration} and \textit{Recalibration SCADANN} with eleven gestures on the dynamic dataset using the first Evaluation Recording as unlabeled data}
\label{comparisonRecalibrationAndRecal_SCADANN}
\resizebox{\linewidth}{!}{%
\begin{tabular}{@{}ccc|cc@{}}
\toprule
 &
  \multicolumn{2}{c|}{Spectrogram ConvNet} &
  \multicolumn{2}{c|}{TSD DNN} \\ \midrule
 &
  Recalibration &
  Recalibration SCADANN  &
  Recalibration &
  Recalibration SCADANN  \\ \midrule
\begin{tabular}[c]{@{}c@{}}Session 0\\ STD\\ H0\\ Cohen's Dz\end{tabular} &
  \begin{tabular}[c]{@{}c@{}}47.81\%\\ 10.94\%\\ 0 (0.00642)\\ N\textbackslash{}A\end{tabular} &
  \textbf{\begin{tabular}[c]{@{}c@{}}49.89\%\\ 11.25\%\\ 0 (0.00642)\\ 0.73\end{tabular}} &
  \begin{tabular}[c]{@{}c@{}}53.08\%\\ 11.48\%\\ 0 (0.01000)\\ N\textbackslash{}A\end{tabular} &
  \textbf{\begin{tabular}[c]{@{}c@{}}55.69\%\\ 12.37\%\\ 0 (0.01000)\\ 0.67\end{tabular}} \\ \midrule
\begin{tabular}[c]{@{}c@{}}Session 1\\ STD\\ H0\\ Cohen's Dz\end{tabular} &
  \begin{tabular}[c]{@{}c@{}}49.54\%\\ 11.28\%\\ 0 (0.00455)\\ N\textbackslash{}A\end{tabular} &
  \textbf{\begin{tabular}[c]{@{}c@{}}53.02\%\\ 11.18\%\\ 0 (0.00455)\\ 0.88\end{tabular}} &
  \begin{tabular}[c]{@{}c@{}}53.51\%\\ 11.98\%\\ 0 (0.00014)\\ N\textbackslash{}A\end{tabular} &
  \textbf{\begin{tabular}[c]{@{}c@{}}58.34\%\\ 11.82\%\\ 0 (0.00014)\\ 1.25\end{tabular}} \\ \midrule
\begin{tabular}[c]{@{}c@{}}Session 2\\ STD\\ H0\\ Cohen's Dz\end{tabular} &
  \begin{tabular}[c]{@{}c@{}}52.18\%\\ 10.66\%\\ 0 (0.00059)\\ N\textbackslash{}A\end{tabular} &
  \textbf{\begin{tabular}[c]{@{}c@{}}55.19\%\\ 10.15\%\\ 0 (0.00059)\\ 1.11\end{tabular}} &
  \begin{tabular}[c]{@{}c@{}}57.18\%\\ 11.12\%\\ 0 (0.00012)\\ N\textbackslash{}A\end{tabular} &
  \textbf{\begin{tabular}[c]{@{}c@{}}60.81\%\\ 10.10\%\\ 0 (0.00012)\\ 0.75\end{tabular}} \\ \bottomrule
\end{tabular}
}
*Wilcoxon signed rank test. Null hypothesis rejected when H0=0 (p$<$0.05).
\end{table}

\section{Discussion}
\label{discussion}

The task of performing adaptation when multiple days have elapsed is especially challenging. As a comparison, on the within-day adaptation task presented in~\cite{self_recalibration_majority_vote}, MV was able to enhance classification accuracy by 10\% on average compared to the No Calibration scheme. Within this work however, the greatest improvement achieved by MV was 3.35\% for the Spectrogram ConvNet and 5.18\% for the TSD DNN. Overall, the best improvement in this paper was 8.47\% and 10.81\% both achieved by SCADANN with the Spectrogram ConvNet and TSD DNN respectively. All three tested domain adversarial algorithms were also able to consistently improve the network's accuracy compared to the No Calibration scheme (the only exception being VADA and Dirt-T for the TSD DNN in Table~\ref{comparison_DA_training_offline_test_online}). When used to adapt to dynamic unsupervised data, some were even able to achieve a higher overall ranking than SCADANN using the Spectrogram ConvNet. Note however, that the improvements they seem to allow is overall lower than when they are applied on image-based dataset such as MNIST and CIFAR~\cite{dirt_T}. Deep domain adversarial algorithms thus seems to be a promising avenue to explore further, by developing adversarial algorithms specifically for the field of sEMG-based gesture recognition. SCADANN could then easily be augmented by these new algorithms to improve performance further.

%This decrease in performance from SCADANN and MV on harder datasets is most likely due to the reduction of the overall classifier's performance. This phenomenon is perhaps best shown by looking at Table~\ref{comparison_DA_training_offline_test_online} and~\ref{comparison_DA_training_online_test_online}, where all algorithms were tested on the same data in both tables. Note how SCADANN was the best ranked adaptation method and MV was the second best on Table~\ref{comparison_DA_training_offline_test_online}, whereas on Table~\ref{comparison_DA_training_online_test_online}, MV performed the worst (alongside AdaBN) and SCADANN was the best ranked method only for the first session (although it consistently obtained the best average accuracy across all methods). 

%When used to adapt to online unsupervised data, they were even able to achieve higher overall accuracy than SCADANN. This decrease of performance from SCADANN and MV on harder datasets is most likely due to the reduction of the overall classifier's performance. This phenomena is perhaps best shown by looking at Table~\ref{comparison_DA_training_offline_test_online} and~\ref{comparison_DA_training_online_test_online}, where all algorithms were tested on the same data in both tables. Note how SCADANN was the best ranked adaptation method and MV was the second best on Table~\ref{comparison_DA_training_offline_test_online}, whereas on Table~\ref{comparison_DA_training_online_test_online} they degenerated into being the worst (with MV being even worst than the No Calibration setting on session 0 and 2). 

In Table~\ref{comparison_DA_training_online_test_online}, it can be seen that the performances of MV dropped substantially compared to the other experiments conducted within this paper. A possible explanation is that this was the first time that MV had to adapt using the Dynamic Dataset data. In other words, instead of adapting to a well defined series of examples grouped by gesture, MV had to contend with a continuous data stream including gesture transitions. In contrast, SCADANN actually performed generally better in Table~\ref{comparison_DA_training_online_test_online} than in Table~\ref{comparison_DA_training_offline_test_online}, which is encouraging as Table~\ref{comparison_DA_training_online_test_online} showcased a more realistic setting for unsupervised recalibration.

It is also important to note that both the general performance and the type of error that the classifier makes can greatly affect the self-calibrating algorithms. While SCADANN partially address the first consideration (by ignoring data that are more likely to be misclassified), the second consideration is harder to address. That is, when the classifier is not only wrong, but is confident in its error and that error spans over a large amount of time, the pseudo-labeling heuristic cannot hope to re-label the segments correctly or even identify this segment of data as problematic. In an effort to address this issue, future works could explore the use of a hybrid IMU/EMG system, as they have been shown to improve gesture recognition accuracy~\cite{emg_plus_imu_classifier, emg_plus_imu_good}. The use of accelerometer data within the field is generally linked with mechanomyogram (MMG), which is strongly associated with EMG signals. Recent works~\cite{evan_imu_emg} however, have shown that, within a human-computer interaction context, accelerometer data can also help recognize different gestures with high accuracy using the positional variance of the different gestures, which is uncharacteristic of MMG. The fusion of these two different modalities could reduce the likelihood of concurrent errors, enabling SCADANN's relabeling heuristic to generate the pseudo-labels more accurately. Note that, using EMG signals alone, SCADANN's relabeling heuristic substantially enhanced the pseudo-labels accuracy compared to the one used with MV. As an example, consider the supervised Recalibrating classifier (with the Spectrogram ConvNet) trained on all the training cycles of the relevant Training Recording and tested on the Evaluation Recording. This classifier achieves an average accuracy of 49.84\% over 544 263 examples. In comparison, the MV relabeling heuristic achieves 54.28\% accuracy over the same number of examples, while the SCADANN relabeling heuristic obtains 61.89\% and keeps 478 958 examples using the 65\% threshold. When using a threshold of 85\%, the accuracy reaches 68.21\% and retains 372 567 examples. SCADANN's improved relabeling accuracy compared to MV is in part due to the look-back feature of the heuristic (when de-activated, SCADANN's relabeling accuracy drops to 65.23\% for the 85\% threshold) and its ability to remove highly uncertain sub-sequences of predictions.
% Within this study, the reason for the 65\% threshold was due to the limited availability of unlabeled data within each session, as removing too many examples might completely erase some gestures. It is suspected that in real application scenarios, where the amount of unlabeled data is not limited, higher thresholds should be preferred as discarding more examples can be afforded and would most likely enhance the performance of the self-calibrated classifier further. 

The results presented in Table~\ref{comparisonRecalibrationAndRecal_SCADANN} are of particular interest as they show that SCADANN actually consistently and significantly improves the classifier's performance over the recalibration scheme. In other words, SCADANN enhance classifier's performance without increasing the training time for the participant. In addition, as SCADANN does not impact the classifier's inference time, SCADANN seems to be an overall net benefit for the classifier's usability.

\subsection{Limitations of the study}
One major limitation of this work is that the participants were not reacting to the different classifiers being tested (instead using the leap-motion based controller) while performing the task from the Evaluation Recording. This limitation is the only difference between the Dynamic dataset and an online dataset. It is important to note that the participants generally became better at performing the requested task over time (see~\cite{VR_long_term_dataset} and Table~\ref{comparisonRecalibrationAndRecal_SCADANN}). The extent to which this improvement can be attributed to the user's adaptation to the leap-motion based controller and how much should be attributed to the participants learning how to complete the task better remains unclear. What is known is that the user's adaptation to the controller substantially affects the real-time control performance of the system~\cite{journal_paper_TL_ulysse, users_adapt_overtime}. If and how much this adaptation changes in relation to the controller use, however, remains an open question to the best of the authors' knowledge. Furthermore, this user adaptation would substantially alter the optimal rate of unsupervised calibration and the acceptable extent of said calibration. These new parameters might be better explored within a reinforcement learning~\cite{reinforcement_learning} framework.

As a direct consequence of not having the adaptation algorithms tested in real-time, another limitation of this work is that the adaptation algorithms were not evaluated using online metrics (e.g. throughput, completion rate, overshoot)~\cite{online_metrics}. To do so would require recording a separate long-term dataset, as extensive as the one used in this work, for each compared technique so that the different adaptive classifier could be used by the participants in real-time. The difficulty of comparing different adaptation algorithms using online metrics was, in fact, the motivation behind the use of the \textit{Long-term 3DC Dataset}~\cite{VR_long_term_dataset} which allows for recording closer to an online setting (compared to offline datasets) without biasing the dataset to a particular EMG-based gesture classification algorithms. Thus, allowing comparison between multiple techniques on a single dataset.

\section{Conclusion}
\label{conclusion}
This paper presents SCADANN, a self-calibrating domain adversarial algorithm for myoelectric control systems. Overall, SCADANN was shown to improve the network's performance compared to the No Calibration setting in all the tested cases and the difference was significant across all experiments except for one single session. In addition, this work tested three widely used, state-of-the-art, unsupervised domain adversarial algorithms on the challenging task of EMG-based self-calibration. These three algorithms were also found to consistently improve the classifier's performance compared to the No Calibration setting. MV~\cite{self_recalibration_majority_vote} and AdaBatch~\cite{AdaBN}, two self-calibrating algorithms designed for EMG-based gesture recognition, were also compared to the three DA algorithms and SCADANN. Overall, SCADANN was shown to consistently obtain the best average accuracy amongst the six unsupervised adaptation methods considered in this work both using offline and dynamic datasets. Given the results shown in this paper and considering that SCADANN has no computational overhead at prediction time, using it to adapt to never-before-seen data is a net benefit both for long-term use but also right after recalibration (as shown in Figure~\ref{train_online_test_online_figure}). 

Future works will focus on implementing SCADANN to update in real-time while in use by participants. The interaction between human and machine adaptation and its impact on self-adaptive algorithms like SCADANN will be investigated by leveraging a reinforcement learning framework.

\appendices

\section{ConvNet's comparison with Handcrafted feature sets}
\label{appendix_comparisons_with_feature_sets}
To better interpret the contributions of this manuscript, it is important to contextualize the ConvNet's classification performances with respect to the state of the art in sEMG-based gesture recognition.

The comparison considers the simple ConvNet employed throughout this work with six high performing feature sets presented in the following subsections. The python implementation of the different feature sets are available on this work's \href{https://github.com/UlysseCoteAllard/LongTermEMG}{repository: (https://github.com/UlysseCoteAllard/LongTermEMG)} and a detailed description of most of the features are given in~\cite{journal_paper_TL_ulysse}. Note that the hyperparameters associated with these feature sets employed the ones recommended in their respective original paper.

\subsection{Hudgin's features}
Hudgin's features~\cite{TD_feature_set} are a set of four features all in the time-domain comprised of: Mean Absolute Value, Zero Crossing, Slope Sign Changes and Waveform Length. As all the features are in the time-domain, this feature set is often referred to (and will be in this work) as TD. TD is arguably the most commonly employed feature set~\cite{overview_emg_interface_2015} and serves as a baseline when comparing different handcrafted feature sets.

\subsection{NinaPro feature set}
The NinaPro feature~\cite{ninaProDB2_DB3, NinaPro_feature_sets_recent} set has been successfully employed on the diverse NinaPro datasets and consist of the concatenation of the TD features alongside Histogram and marginal Discrete Wavelet Transform.

\subsection{SampEn pipeline}
The SampEn pipeline~\cite{Sampen_pipeline_feature_set} consists of Sample Entropy, Cepstral Coefficients, Root Mean Square and Waveform Length. This feature set was found to be the best combination of features amongst the 50 considered in the original work (brute-force search).

\subsection{LSF9}
LSF9~\cite{LSF9_early, LSF9_recent} is a newly proposed feature set which was originally developed specifically for low sampling rate recording devices (200\textit{Hz}). Nevertheless, this feature set also offers exceptional performance on higher sampling rate datasets. LSF9 consists of: L-scale, Maximum Fractal Length, Mean Value of the Square Root, Willison Amplitude, Zero Crossing, Root Mean Square, Integrated Absolute Value, Difference Absolute Standard Deviation Value and Variance.

\subsection{TDPSD}
TDPSD~\cite{TDPSD_early, TDPSD_recent} proposes to consider the EMG signal alongside their nonlinear cepstral representation. Then, one vector per representation is created by computing the: Root squared zero, second and fourth moments as well as Sparseness, Irregularity Factor and the Waveform Length Ratio. The final vector used for classification is obtained from the cosine similarity of the two previous vectors. The interested reader is encouraged to consult~\cite{TDPSD_early} for a detailed description of this feature set.

\subsection{TSD}
TSD~\cite{TSD} represents the evolution of TDPSD. The idea of leveraging the cosine similarity between two vectors of the same features computed from different representation of the signal remain. However, the features have been updated and now consist of: the Root squared zero, second and fourth moments as well as the Sparseness, Irregularity Factor, Coefficient of Variation and the Teager-Kaiser energy operator. Most importantly, this feature set not only considers the similarities between the signal of a particular channel and its nonlinear transformation but also considers these similarities across channels. The interested reader is encouraged to consult~\cite{TSD} for a detailed description of this feature set.

\subsection{Dataset and Classifier}
A standard Linear Discriminant Analysis~\cite{overview_emg_interface_2015} is selected for classification as it is widely employed in the field and is a computationally and time efficient classification technique both at training and prediction time, while still achieving high classification accuracy~\cite{overview_emg_interface_2015, journal_paper_TL_ulysse, Sampen_pipeline_feature_set}.

The Long-term 3DC Dataset is employed for comparison. For each Training Recording of each participant (20 participants $\times$ 3 sessions). The first two cycles are employed for training, while the last cycle is reserved for testing (total of 60 train/test per method). The comparison is done for both the seven and eleven gestures considered in this work. The ConvNet's architecture and hyperparameters are exactly as described in Section~\ref{spectrogram_architecture}. The LDA implementation is from scikit-learn~\cite{scikit_learn} with its defaults parameters.

\subsection{Comparison of results}

Table~\ref{comparison_feature_sets_table} presents the comparison between the ConvNet and the six feature sets.

\begin{table}[!htbp]
\caption{Comparison between the ConvNet employed in this work and Handcrafted feature sets}
\label{comparison_feature_sets_table}
\resizebox{\linewidth}{!}{%
\begin{tabular}{cccccccc}
\hline
 & ConvNet & TD & NinaPro & \begin{tabular}[c]{@{}c@{}}SampEn \\ Pipeline\end{tabular} & LSF9 & TDPSD & TSD \\ \hline
\begin{tabular}[c]{@{}c@{}}7 Gestures\\ STD\\ Friedman Rank\\ H0\end{tabular} & \begin{tabular}[c]{@{}c@{}}93.13\%\\ 6.44\%\\ 3.50\\ N\textbackslash{}A\end{tabular} & \begin{tabular}[c]{@{}c@{}}89.18\%\\ 8.34\%\\ 5.86\\ 0 (\textless{}0.00001)\end{tabular} & \begin{tabular}[c]{@{}c@{}}89.48\%\\ 7.87\%\\ 5.61\\ 0 (\textless{}0.00001)\end{tabular} & \begin{tabular}[c]{@{}c@{}}91.03\%\\ 7.48\%\\ 4.42\\ 0 (0.04023)\end{tabular} & \begin{tabular}[c]{@{}c@{}}94.45\%\\ 5.89\%\\ 2.51\\ 0 (0.03578)\end{tabular} & \begin{tabular}[c]{@{}c@{}}92.67\%\\ 6.26\%\\ 3.98\\ 1\end{tabular} & \textbf{\begin{tabular}[c]{@{}c@{}}95.01\%\\ 5.47\%\\ 2.13\\ 0 (0.00196)\end{tabular}} \\ \hline
\begin{tabular}[c]{@{}c@{}}11 Gestures\\ STD\\ Friedman Rank\\ H0\end{tabular} & \begin{tabular}[c]{@{}c@{}}85.42\%\\ 9.69\%\\ 4.07\\ N\textbackslash{}A\end{tabular} & \begin{tabular}[c]{@{}c@{}}81.11\%\\ 9.97\%\\ 5.70\\ 0 (0.00017)\end{tabular} & \begin{tabular}[c]{@{}c@{}}81.32\%\\ 9.80\%\\ 5.52\\ 0 (0.00095)\end{tabular} & \begin{tabular}[c]{@{}c@{}}83.57\%\\ 9.71\%\\ 4.41\\ 1\end{tabular} & \begin{tabular}[c]{@{}c@{}}87.94\%\\ 9.26\%\\ 2.69\\ 0 (0.0.00147)\end{tabular} & \begin{tabular}[c]{@{}c@{}}84.86\%\\ 9.61\%\\ 4.01\\ 1\end{tabular} & \textbf{\begin{tabular}[c]{@{}c@{}}91.03\%\\ 8.73\%\\ 1.61\\ 0 (\textless{}0.00001)\end{tabular}} \\ \hline
\end{tabular}%
}%
\end{table}

When testing on the Evaluation Recording, the ConvNet obtained an average accuracy of 49.84\%$\pm$10.93\%, while TD obtained 48.90\%$\pm$10.80\%, TDPSD obtained 50.55\%$\pm$10.89\% and TSD obtained 56.50\%$\pm$11.27\%. The comparison shows that despite the simplicity of the ConvNet used in this work, it performs almost identically to TDPSD on average and similarly to the five other feature sets considered.

\section{Pseudo-labeling Heuristic}
\label{pseudo_labels_appendix}
\begin{algorithm*}[!htbp]
\caption{Pseudo-labeling Heuristic}
\label{pseudo_labeling_heuristic}
\begin{algorithmic}[1]
\Procedure{GeneratePseudoLabels}{\textit{unstable\_len}, \textit{threshold\_stable}, \textit{max\_len\_unstable}, \textit{max\_look\_back}, \textit{threshold\_derivative}}
\State \textit{pseudo\_labels} $\gets$ empty array
\State \textit{arr\_preds} $\gets$ network's predictions
\State \textit{arr\_net\_out} $\gets$ network's softmax output 
\State \textit{begin\_arr} $\gets$ The \textit{unstable\_len} first elements of \textit{arr\_net\_out}
\State \textit{stable} $\gets$ TRUE
\textit{arr\_unstable\_output} $gets$ empty array
\State \textit{current\_class} $\gets$ The label associated with the output neuron with the highest median value in \textit{begin\_arr}
\For{$i$ from 0..\textit{arr\_preds} length}      

        \If{\textit{current\_class} different than \textit{arr\_preds[i]} AND \textit{stable} TRUE}
        \State \textit{stable} $\gets$FALSE
        \State \textit{first\_index\_unstable} $\gets$ i
        \State \textit{arr\_unstable\_output} $\gets$ empty array
        \EndIf
        \If{\textit{stable} is FALSE}
            \State APPEND \textit{arr\_net\_out} to \textit{arr\_unstable\_output}
            \If{length of \textit{arr\_unstable\_output} is greater than \textit{unstable\_len}}
                \State REMOVE the oldest element of \textit{arr\_unstable\_output}
            \EndIf
            
            \If{length of \textit{arr\_unstable\_output} is greater or equal to \textit{unstable\_len}}
                \State \textit{arr\_median} $\gets$ The median value in \textit{arr\_unstable\_output} for each gesture
                \State \textit{arr\_percentage\_medians} $\gets$ \textit{arr\_median} / the sum of \textit{arr\_median}
                \State \textit{gesture\_found} $\gets$ The label associated with the gesture with the highest median percentage from \textit{arr\_percentage\_medians}
                
                \If{\textit{arr\_percentage\_medians[gesture\_found]} greater than \textit{threshold\_stable}}
                    \State \textit{stable} $\gets$ TRUE
                    \If{\textit{current\_class} is \textit{gesture\_found} AND The time within instability is less than \textit{max\_len\_unstable}}
                        \State Add the predictions which occurred during the unstable time to \textit{pseudo\_labels} with the \textit{gesture\_found}
                        
                    \ElsIf{\textit{current\_class} is different than \textit{gesture\_found} AND The time within instability is less than \textit{max\_len\_unstable}}
                        \State \textit{index\_start\_change} $\gets$ GetIndexStartChange(\textit{arr\_net\_out}, \textit{first\_index\_unstable}, \textit{max\_look\_back})
                        \State Add the predictions which occurred during the unstable time to \textit{pseudo\_labels} with the \textit{gesture\_found} label
                        \State Re-label the predictions from \textit{pseudo\_labels} starting at \textit{index\_start\_change} with the \textit{gesture\_found} label
                    \EndIf
                    \State \textit{current\_class} $\gets$ \textit{gesture\_found}
                    \State \textit{arr\_unstable\_output} $\gets$ empty array
                \EndIf
            \EndIf
        \Else
            \State Add current prediction to \textit{pseudo\_labels} with the \textit{current\_class} label
        \EndIf
\EndFor
\Return \textit{pseudo\_labels}
\EndProcedure
\end{algorithmic}
\end{algorithm*}
\vspace{-2cm}
\begin{algorithm*}[!htbp]
\caption{Find index start of transition heuristic}
\begin{algorithmic}[1]
\Procedure{GetIndexStartChange}{\textit{arr\_net\_out}, \textit{first\_index\_unstable}, \textit{max\_look\_back}, \textit{threshold\_derivative}}
\State \textit{data\_uncertain} $\gets$ Populate the array with the elements from \textit{arr\_net\_out} starting from the \textit{first\_index\_unstable}-\textit{max\_look\_back} index to the \textit{first\_index\_unstable} index
\State \textit{discrete\_entropy\_derivative} $\gets$ Calculate the entropy for each element of \textit{data\_uncertain} and then create an array with their derivatives. 
\State \textit{index\_transition\_start} $\gets$ 0
\For{$i$ from 0..\textit{data\_uncertain} length}        
    \If{\textit{discrete\_entropy\_derivative[i]} greater than \textit{threshold\_derivative}}
        \State \textit{index\_transition\_start} $\gets i$
        \State Get out of the loop
    \EndIf
\EndFor
\Return \textit{first\_index\_unstable} + \textit{index\_transition\_start}
\EndProcedure
\end{algorithmic}
\end{algorithm*}
\clearpage
\bibliographystyle{IEEEtran}
\bibliography{main}

\clearpage

\begin{IEEEbiography}[{\includegraphics[width=1in,height=1.25in,clip,keepaspectratio]{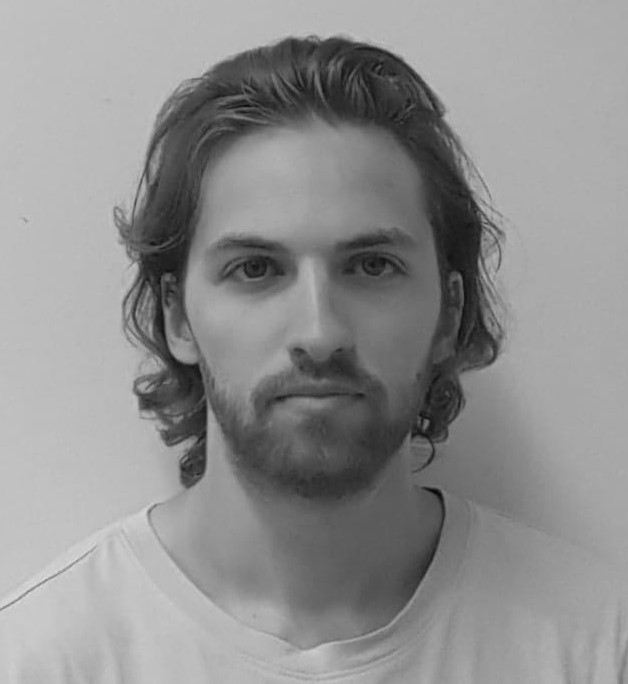}}]%
{Ulysse Côté-Allard} received the Ph.D. degree in electrical engineering from Université Laval, Québec, QC, Canada, in 2020. He is currently completing a Postdoctoral fellow at the University of Oslo, Oslo, Norway with the Robotics and Intelligent Systems research group. 

His main research interests include rehabilitation engineering, biosignal-based control, and human-robot interaction.
Dr. Côté-Allard is the recipient of the Best Paper Award from the IEEE Systems, Man, and Cybernetics conference. 
\end{IEEEbiography}

\begin{IEEEbiography}[{\includegraphics[width=1in,height=1.25in,clip,keepaspectratio]{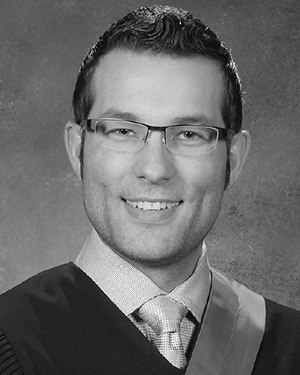}}]%
{Gabriel Gagnon-Turcotte} Gabriel Gagnon-Turcotte (S’15) received the Ph.D. degree in electrical engineering from Laval University, Quebec city, QC, Canada, in 2019. He is currently working as a full time researcher in electrical engineering at the Biomedical Microsystems Laboratory, Laval University. His main research interests are neural compression algorithms, wireless implantable biomedical systems, mixed-signal/analog IC design, system-level digital design, and VLSI signal processing. He has been a recipient of several awards, including the BioCAS’15 Best Paper Award (runner up), the Canadian Governor General's Academic Gold Medal, and the Brian L. Barge Microsystems Integration Award.
\end{IEEEbiography}
\begin{IEEEbiography}[{\includegraphics[width=1in,height=1.25in,clip,keepaspectratio]{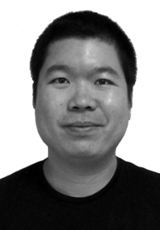}}]%
{Angkoon Phinyomark} Angkoon Phinyomark (M’09) was born in Thailand in 1986. He received the B.Eng. (Hons.) degree in computer engineering and the Ph.D. degree in electrical engineering from the Prince of Songkla University, Songkhla, Thailand, in 2008 and 2012, respectively.
He was a Post-Doctoral Research Fellow with the GIPSA Laboratory and the LIG Laboratory, University Joseph Fourier, Grenoble, France, from 2012 to 2013. Since 2013, he has been a Post-Doctoral Research Fellow with the Human Performance Laboratory, University of Calgary, Calgary, AB, Canada. His current research interests include biomedical signal processing and pattern recognition notably electromyography signal, dimensionality reduction, machine learning, human movement, biomechanics, running gait analysis, musculoskeletal injury, wavelet analysis, and fractal analysis. He is currently a Senior Research Scientist at Institute of Biomedical Engineering at the University of New Brunswick. 
\end{IEEEbiography}

\begin{IEEEbiography}[{\includegraphics[width=1in,height=1.25in,clip,keepaspectratio]{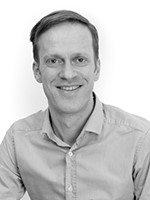}}]%
{Kyrre Glette} received his Ph.D. in Computer Science from the University of Oslo, Norway in 2008. 
He is currently an Associate Professor at the Robotics and Intelligent Systems group (ROBIN), Dept. of Informatics, and a PI at the RITMO Centre for Interdisciplinary Studies in Rhythm, Time and Motion, University of Oslo. Glette has experience in artificial intelligence and adaptive systems, digital design, rapid prototyping, music technology, and robotics. His current research interests include algorithms for automatic adaptation and design of behaviors and shapes for robotic systems, transferring behaviors between simulation and reality, and robotic collectives.
\end{IEEEbiography}

\begin{IEEEbiography}[{\includegraphics[width=1in,height=1.25in,clip,keepaspectratio]{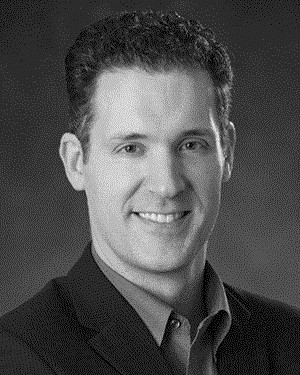}}]%
{Erik J. Scheme}  received the B.Sc., M.Sc., and Ph.D. degrees in electrical engineering from the University of New Brunswick (UNB), Fredericton, NB, Canada, in 2003, 2005, and 2013, respectively.
He is an Assistant Professor with the Department of Electrical and Computer Engineering, UNB, the New Brunswick Innovation Research Chair in Medical Technologies with the Institute of Biomedical Engineering, an Adjunct Professor in Medical Education with Dalhousie Medicine New Brunswick, and the Director of the Health Technologies Innovation Laboratory with UNB. His research interests include human–machine interfaces, biological signal processing, and diagnostics and predictive analytics for healthcare applications.
Dr. Scheme is a Registered Member of the Association of Professional Engineers and Geoscientists of New Brunswick.
\end{IEEEbiography}

\begin{IEEEbiography}[{\includegraphics[width=1in,height=1.25in,clip,keepaspectratio]{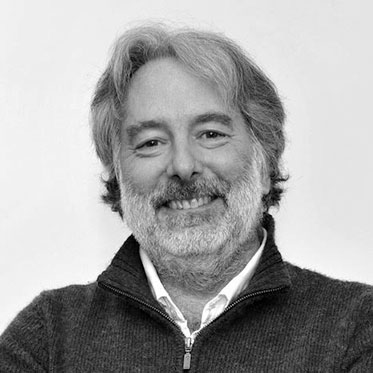}}]%
{François Laviolette} François Laviolette received his Ph.D. in mathematics from Université de Montréal in 1997. His thesis solved a long-standing conjecture (60 years old) on graph theory and was among the seven finalists of the 1998 Council of Graduate Schools/University Microfilms International Distinguished Dissertation Award of Washington, in the category Mathematics-Physic-Engineering. He then moved to Université Laval, where he works on Probabilistic Verification of Systems, Bio-Informatics, and Machine Learning, with a particular interest in PAC-Bayesian analysis.
\end{IEEEbiography}

\begin{IEEEbiography}[{\includegraphics[width=1in,height=1.25in,clip,keepaspectratio]{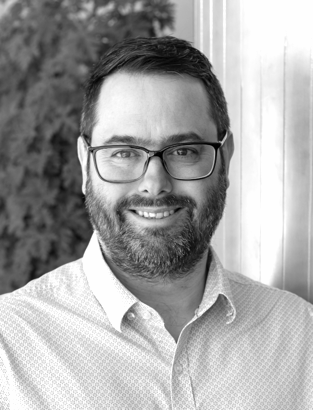}}]%
{Benoit Gosselin} (S’02–M’08) obtained the Ph.D. degree in Electrical Eng. from École Polytechnique de Montréal in 2009, and he was an NSERC Postdoctoral Fellow at the Georgia Institute of Technology in 2010. He is currently a Full Professor at the Depart. of ECE at Université Laval, where he holds the Canada Research Chair in Smart Biomedical Microsystems. His research interests include wireless microsystems for brain computer interfaces, analog/mixed-mode and RF integrated circuits for neural engineering, interface circuits of implantable sensors/actuators and point-of-care diagnostic microsystems for personalized healthcare. Dr Gosselin is an Associate Editor of the IEEE Transactions on Biomedical Circuits and Systems and he is Chair and Founder of the IEEE CAS/EMB Quebec Chapter (2015 Best New Chapter Award). He served on the committees of several int’l IEEE conferences including BIOCAS, NEWCAS, EMBC, LSC and ISCAS. Currently, he is Program Chair of EMBC 2020, the first virtual EMBC in response to the COVID-19 pandemic. His significant contribution to biomedical microsystems research led to the commercialization of the first wireless microelectronic platform to perform optogenetics and electrophysiology in parallel with his partner Doric Lenses Inc. He is Fellow of the Canadian Academy of Engineering, and he has received several awards, including the prestigious NSERC Brockhouse Canada Prize, and the Prix Génie Innovation of the Quebec professional engineering association OIQ.
\end{IEEEbiography}

\end{document}